\documentclass[aps,showpacs]{revtex4}

\usepackage{bm}
\usepackage{epsfig}
\usepackage{amsmath}

\def\be{\begin{equation}}
\def\lan{\left\langle}
\def\ran{\right\rangle}
\def\ee{\end{equation}}
\def\barr{\begin{array}}
\def\earr{\end{array}}

\def\nn8{\nonumber\\[15pt]}
\def\l{\left}
\def\r{\right}
\def\dis{\displaystyle}
\def\ed{\end{document}}
\def\cg{\cal{G}}
\def\ce{\cal{E}}
\def\dg{\dagger}
\begin{document}

\title{Chaos and localization in the wavefunctions of 
       complex atoms NdI, PmI and SmI}

\author{Dilip Angom and V. K. B. Kota}
\affiliation{ Physical Research Laboratory, 
              Navarangpura, Ahmedabad - 380 009}

\begin{abstract}

Wavefunctions of complex lanthanide atoms NdI, PmI and SmI, obtained via
multi-configuration Dirac-Fock method, are analyzed for density of states in
terms of partial densities, strength functions ($F_k(E)$), number of
principal components ($\xi_2(E)$) and occupancies ($\lan n_\alpha
\ran^E$) of single particle orbits using embedded Gaussian orthogonal
ensemble of one plus two-body random matrix ensembles [EGOE(1+2)]. It is
seen that density of states are in general multi-modal,  $F_k(E)$'s exhibit
variations as function of the basis states energy and $\xi_2(E)$'s show
structures arising from localized states.  The sources of these departures
from EGOE(1+2) are investigated by examining the partial densities, 
correlations between $F_k(E)$, $\xi_2(E)$ and $\lan n_\alpha \ran^E$ and also 
by studying the structure of the Hamiltonian matrices. These
studies point out the operation of EGOE(1+2) but at the same time  suggest
that weak admixing between well separated configurations should be
incorporated into EGOE(1+2) for more quantitative description of chaos and
localization in NdI, PmI and SmI. 

\end{abstract}



\maketitle
\date{}


\section{Introduction}
\label{introduction}

Lanthanide atoms exhibit complicated configuration mixing and have complex
spectra \cite{martin} as consequence of partially filled high  angular
momentum $4f$ and $5d$ valence shells. The general form of the ground state 
configuration  is $[{\rm Xe}]6s^24f^m5d^n$, where $m$ and $n$ are the shell
occupancies. Across the period $m$ increases from 1 to 14, whereas  $n$ is 1
for Ce I and Gd I   which are located at the beginning and middle of the
period and $n$ is 0 for the remaining lanthanide atoms. Along  the period the
number of configurations and mixing increases as $4f$ shell occupancy
increases, and reaches a  maximum around the middle. Lanthanide atoms are
appropriate to  study complexities emerging from two-body interactions and a
comparative study along the period could  provide insights to the
implications of  increasing particle number to the nature of the complexity. 

Detailed  studies of Ce I  wave-function structure, first carried out by
Flambaum  and collaborators \cite{Fl-94,Fl-99}  showed the smoothed (with
energy) strength functions ($F_k(E)$) and number of principle components 
(NPC or $\xi_2(E)$), which measure chaoticity, can be understood in terms of
banded random matrices (BRM) with the  local level and strength fluctuations
following the predictions of Wigner's Gaussian orthogonal ensemble (GOE) of
random matrices.  Later studies of Cummings and collaborators on Pr I  (and
also Ce I)  \cite{Cu-01}, the element  next to Ce I in the lanthanide series,
showed results in agreement with Flambaum et al studies of CeI.  Flambaum
and  collaborators further developed, using the BW form for
$F_k(E)$ and NPC in terms of the BW spreading widths which are a result of
many body chaos, a statistical  theory for the distribution of  occupation
numbers \cite{Fl-97,Fl-97b}, and also for electromagnetic transition
strengths in atoms  with complex (or chaotic) spectra \cite{Fl-96}. Applying
this statistical  spectroscopy for complex states, energy-averaged cross
sections are obtained in terms of sums over single electron orbitals and
using this Flambaum et al \cite{Fl-02} and Gribakin et al \cite{Gr-03} 
explained the observed, and quite puzzling, low-energy electron recombination 
rate of $Au^{25+}$.  

Following  the successful applications in nuclei
\cite{Br-81,Ko-01,Nu-03,Ko-03}, mesoscopic physics \cite{Ja-01},  in the
context of quantum computers \cite{Qc-00} and as suggested more  recently for
example in the dynamics of  cold atoms \cite{Be-04}, a better random matrix
hypothesis for atoms is to consider the interaction (after subtracting the
single particle field generated by the Coulomb Hamiltonian 
see Eq. (\ref{hdc}) ahead
for the Hamiltonian) to be random. This gives rise to embedded GOE of one
plus  two-body interactions [EGOE(1+2)]. See Sect. III ahead for details of
EGOE(1+2) and here it suffices to say that the ensemble is $\{H\} = h(1) +
\lambda \{V(2)\}$ where $h(1)$ is the mean-field one-body part and $\{V(2)\}$
is a GOE in 2-particle space with $\lambda$ the interaction strength. An
extreme limit, with strong interaction ($\lambda \rightarrow \infty$),  of
EGOE(1+2) is two body random matrix ensemble (TBRE). Recently we demonstrated
that Sm I \cite{Dk-03}, which has three more active electrons compared to Pr
I, carries TBRE signatures which include Gaussian state densities, strength
functions having Gaussian form, NPC as a function of energy is of Gaussian
form, local level fluctuations follow GOE etc. Going beyond this limiting
situation, EGOE(1+2) with increasing interaction strength (equivalently with
increasing particle number if the interaction is fixed), strength functions
exhibit Breit-Wigner to Gaussian transition. It is important to stress that
EGOE(1+2) at the weak limit ($\lambda$ small) shares features with BRM, for
example the BW form for strength functions (this explains partly the
applicability of BRM for CeI and PrI). In a recent paper \cite{Agk-03}, an
interpolating function, analogous to the Brody distribution for nearest
neighbor spacing distribution, representing the  BW to Gaussian transition
of the strength function is reported (see Eq. (\ref{fitfun}) ahead) and it is 
used to understand this transition  in rare earth atoms (at the spectrum's 
center) as we go from CeI to SmI with valence electrons changing from 4 to 8. 
In addition, it is noticed in NdI, PmI and SmI  studies that the eigenfunctions
exhibit localization properties.  The presence of localized states is also
evident in the study of Ce I and Pr I by Cummings and collaborators
\cite{Cu-01}.

As shown in  \cite{bauche-90}, the number of allowed $4f^m6s^2 - 4f^m6s6p$
and  $4f^m - 4f^{m-1}5d$ dipole transitions of the rare earth atoms is large 
( for example, in Pr I   the number of allowed $4f^36s^2 - 4f^36s6p$ dipole
transitions is 7402) and it is difficult to study the transitions
individually. However, statistical  spectroscopy, based on TBRE and many-body
chaos, provides a means to understand the smoothed part of these  transitions
(with fluctuations following GOE). In this paper we consider NdI, PmI and SmI
with 6,7 and 8 valence electrons and examine in some detail: (i) the extent
to which EGOE(1+2) applies for these atoms; (ii) localization properties of
wavefunctions; (iii) the structure of the Hamiltonian matrix. For (i) and
(ii), wavefunctions are analyzed in terms of density of states as sum of
partial densities, strength functions, NPC and occupancies of single particle
orbits and correlations between them. These results form Section IV of the
paper. Section V discusses the structure of the Hamiltonian matrices and the
modifications to be incorporated in EGOE(1+2) for a more quantitative
description of chaos and localization in lanthanide atoms. For completeness,
Section II gives the method of atomic structure calculations and Section III
gives the definition and a brief discussion of some of the basic results for
EGOE(1+2). Finally Section V gives conclusions and future outlook.


\section{Method of atomic calculations}
\subsection{Multi-Configuration Dirac-Fock method}
\label{mcdf}

The earlier works on lanthanide atoms \cite{angom-01} have shown that the 
multiconfiguration Dirac-Fock ( MCDF) method is suitable for studying the
structure  and calculating properties of these atoms. The method is the 
relativistic equivalent of multiconfiguration  Hartree-Fock (MCHF) 
\cite{froese} used
extensively in  non-relativistic atomic calculations.  For a detailed
description of MCDF see ref \cite{grant} and several groups have implemented
it to study atom and molecules \cite{parpia}. For completeness and
continuity  important features of the method are described in this section.
The essence of MCDF is the calculation of orbitals in a mean field arising
from a linear combination of configurations. Then, the single particle states
are the eigenstates of variationally calculated single-electron Schr\"odinger
equations and
these are represented as $|n\kappa\rangle $, where $n$ is the principal
quantum number. The quantum number   $\kappa = \mp(l\pm s) + 1/2$ is similar
to the total angular momentum quantum   number $j = l\pm s$, however unlike
$j$ it is unique for each orbital symmetry. The  single electron wavefunction
is the two-component Dirac spinor
\begin{equation}
   \psi_{n\kappa}(\bm{r}) =\langle\bm{r}|n\kappa\rangle = 
   \frac{1}{r} \left ( \begin{array}{c}
                        P_{n\kappa}(r)  \chi_{\kappa m} (\theta, \phi) \\
                        iQ_{n\kappa}(r) \chi_{-\kappa m}(\theta, \phi)
                       \end{array}
               \right )
  \label{orbtl}
\end{equation}
where $P_{n\kappa}(r)$ and $Q_{n\kappa}(r)$ are large and small component
radial functions respectively, and $\chi_{\kappa m} (\theta, \phi)$ and
$\chi_{-\kappa m}(\theta, \phi)$ are the spinor spherical harmonics in the
$lsj$ coupling scheme. The many electron basis functions of the method are
the  configuration state functions (CSF's) \cite{froese}. Each CSF is
represented as  $|\gamma PJM\rangle$, where $P$, $J$ and $M$ are the parity,
total angular  momentum and magnetic quantum numbers, and $\gamma$ is a
quantum number to  identify each CSF uniquely.  The CSFs are constructed in
two steps: first  couple the identical electrons in each sub-shell $n_i\kappa
_i$ to give $X_i$,   and second couple $X_i$ to obtain $J$. The construction
of the CSFs are better described in second quantized notations, let
$a^{\dagger}_{njm}$ and $a_{njm}$ be  creation and annihilation
operators of the single electron states. Then, the $a^{\dagger}_{njm}$ 
operators form a complete set of spherical tensor  of rank $j$. However, for 
the annihilation operators 
\begin{equation}
    \widetilde{a}_{j m} = ( -1)^{j - m}a_{j, -m}
  \label{anni}
\end{equation}
forms a complete set and not $a_{njm}$. The operators satisfy the
modified anticommutation relation
\begin{equation}
   [\widetilde{a}_{nj}\times a^{\dagger}_{n'j'}]^{JM} + (-1)^{j+j'-J}
   [a^{\dagger}_{n'j'}\times\widetilde{a}_{nj}]^{JM} = \sqrt{2j + 1}
   \delta_{jj'}\delta_{J0}\delta_{M0}\delta_{nn'},
  \label{comm}
\end{equation}
where $[\widetilde{a}_{nj}\times a^{\dagger}_{n'j'}]^{JM}$ represents 
tensor coupling of the operators to rank $J$ and component $M$. In absence 
of external magnetic field, like in the present calculations, the CSFs are 
degenerate in $M$ and each CSF can be identified without $M$ as 
$|\gamma PJ\rangle$. A CSF $|\gamma PJ
\rangle$ having $p$ sub-shells is  created from the vacuum state
\begin{equation}
  |\gamma PJ\rangle  =  [(a^{\dagger}_p)^{q_p}, \nu _p X_p
                        [(a^{\dagger}_{p - 1})^{q_{p - 1}}, \nu _{p - 1}
                        X_{p - 1}[ \ldots 
                        [(a^{\dagger}_1)^{q_1}, \nu _1, X_1]^{J_1} ]
                        \ldots ]^{J_{p - 1}}]^J|0\rangle
\end{equation}
where $q_i$ is the number of electrons in the $i^{\rm th}$ sub-shell and 
$\nu _i$ is the seniority quantum number, which identify identical 
sub-shell total angular momentum $X_i$ uniquely. It is also possible to use 
an ordering index instead of $\nu$ for details; see ref \cite{judd}. 

An appropriate relativistic Hamiltonian to describe high $Z$ atoms like 
lanthanides is Dirac-Coulomb Hamiltonian $H^{\rm DC}$, which includes only 
the electrostatic interactions. For $N$ electron atom
\begin{equation}
   H^{\rm DC}  =  \sum_{i= 1}^{N} c\bm{\alpha}_i\cdot
                  \bm{p}_i + c^2(\beta_i - 1) - \frac{Z (\bm{r}_i)}{r_i} + 
                  \sum_{i=1}^{N}\sum_{j=i+1}^N\frac{1}{|\bm{r}_i - 
		  \bm{r}_j| } 
  \label{hdc}
\end{equation}
where $\bm{\alpha}_i$ and $\beta_i$ are Dirac matrices, $\bm{p}_i$ is the
electron momentum, $Z (\bm{r}_i)$  is the nuclear-charge at
$\bm{r}$ and $N$ is the number of electrons. The first two terms are
single electron Dirac Hamiltonian, the third and last terms  are
electron-nucleus and electron-electron Coulomb interactions respectively. 
The $H^{\rm DC}$  is diagonal in the total angular momentum 
$J$ states $|\Gamma P J\rangle$, the atomic state functions (ASF). Here,
$\Gamma$ is a quantum number to identify each ASF uniquely and parity
$P$ is a good quantum number as $H^{\rm DC}$ is invariant under parity 
transformation. The ASFs are linear combination of CSFs 
$|\gamma_r P J\rangle$
\begin{equation}
   |\Gamma P J\rangle = \sum_{\gamma}c_{r\Gamma} |\gamma_r P J\rangle , 
  \label{asf}
\end{equation}
The ASFs satisfy the Sch\"odinger equation
\begin{equation}
    H^{\rm DC}|\Gamma P J\rangle = E_{\Gamma}|\Gamma P J\rangle 
  \label{scheq}
\end{equation}
where $E_{\Gamma}$ is the eigenvalue. 

The starting point of the MCDF method is the variational optimization of an 
energy functional defined in terms of $H^{\rm DC}$ and Lagrange multipliers
with respect to one or more ASFs. The parameters of variational optimization
are the coefficients $c_{\gamma r}$  and orbitals $\psi_{n\kappa}(\bm{r})$.
In general, the energy  functional is extremized with respect to the ground 
state. Another class of MCDF calculations is the extended optimal level 
(EOL), where the energy  functional is extremized using a set of ASFs. We
use the later method in our calculations and the advantages over the general
method is explained in the next subsection. The energy functional of MCDF-EOL 
calculation is
\begin{equation}
  W^{\rm DC} = \sum_{r,s}^{n_c}d_{rs}H_{rs}^{\rm DC} + \sum_{a = 1}^{n_w}
               \sum_{r = 1}^{n_c}d_{rr}q(a)\epsilon_a + 
               \sum_{a = 1}^{n_w - 1}\sum_{b = a + 1}^{n_w}\delta_{ab}
               \epsilon_{ab}N(ab) , 
  \label{w_dc}
\end{equation}
where $d_{rs}$ and $d_{rr}$ are the weight factors, $H_{rs}^{\rm DC}$ are the
matrix element of $H^{\rm DC}$ between the CSFs $|\gamma_rPJ\rangle$ and
$|\gamma_sPJ\rangle$ and $n_w$ is the number of the orbitals. The quantity
$N(ab)$ is the overlap integral between the $a^{\rm th}$ and $b^{\rm th}$
orbitals and $\epsilon_a$ and $\epsilon_{ab}$ are the Lagrange multipliers to
enforce orthonormality between orbitals of the same symmetry but different
principal quantum numbers. The weight factors can be chosen in several ways, 
in EOL calculation
\begin{equation}
   d_{rs} = \frac{1}{n_L}\sum_i^{n_L}c_{\Gamma_i}^r c_{\Gamma_i}^s
 \label{d_rs}
\end{equation}
where $n_L$ is number of ASFs. Thus, atomic structure and property 
calculations are carried out using a hierarchy of eigenstates: orbitals,
configuration state functions and atomic state functions.


\subsection{Orbital calculation and configuration interaction}
\label{orb-csf}

The calculation of an appropriate set of orbitals $\{ \psi(\bm{r}) \}$ is 
crucial in atomic structure and properties calculations. In the present work 
we calculate the orbitals using the MCDF method described in the previous 
subsection. This method can include strong configuration  mixing, however the 
orbitals generated are state specific, this is undesirable for  statistical 
studies involving several excited states. To strike a balance between these two 
features of MCDF method, orbitals are generated within  selected CSF space and 
the energy functional is optimized with respect to several states using 
MCDF-EOL method. The orbitals generated are less  state specific as these are
calculated in the potential of several states self-consistently. The orbital 
set of each atom are generated in a series of calculations described in this
section.

The calculations are relativistic, however for compact notation we define 
configurations in non-relativistic notations. The Xe like core is considered 
as reference state $|\Phi_0\rangle$ and in non-relativistic notations
\begin{equation}
   |\Phi_0\rangle = [(a^{\dagger}_{5p})^{6}[(a^{\dagger}_{5s})^{2}
                    [ \ldots [(a^{\dagger}_{1s})^{2}]] 
                    \ldots ]]| 0\rangle.
  \label{phi0}
\end{equation}
where $|0\rangle$ is the vacuum state. The intermediate angular momenta $J_i$ 
are zero as all the sub-shells are 
completely filled. The orbitals of the ground state configuration 
$[{\rm Xe}]6s^24f^m5d^n$ are then generated treating $6s$, $4f$ and $5d$ 
orbitals as active valence shells. The calculation is in the CSF manifold of all
single and double replacements among the active valence sub-shells of the 
ground configuration.  The CSFs considered have the form
\begin{equation}
   |\gamma_iP_iJ_i\rangle = [(a^{\dagger}_{5d})^{p_i},\nu_{5d}X_{5d}
                            [(a^{\dagger}_{6s})^{q_i},\nu_{6s}X_{6s}
                            [(a^{\dagger}_{4f})^{r_i},\nu_{4f}X_{4f}
                            ]^{J_{1i}}]^{J_{2i}}]^{J_i}|\Phi_0\rangle,
  \label{grnd}
\end{equation}
where $p_i$, $q_i$ and $r_i$ are the sub-shell occupancies and satisfy the 
condition $p_i + q_i +
r_i = 2 + m + n$. Further, the conditions $m\geq r_i \geq m-2$,  $2\geq q_i \geq 0$
and $n\geq p_i \geq n-2$ are imposed to select single and  double
replacement CSFs within the ground configuration. Then, a MCDF-EOL
calculation of the the lowest multiplet levels is carried out. Valence
orbital  $6p$ is generated in another calculation with frozen core
approximation, where  the orbitals of the ground configuration generated in
the previous calculation are held fixed. The CSF space of the calculation
consists of single excitation  from the ground configuration to $6p$. The
orbital $5d$ is generated in a similar calculation if it is unoccupied ( $n = 0$) in the ground
configuration. Thus, in  relativistic notations, the orbital set consists of
$(1-6)s_{1/2}$,  $(2-6)p_{1/2, 3/2}$, $(3-5)d_{3/2, 5/2}$ and $4f_{5/2,
7/2}$.  Out of these  $6s_{1/2}$, $4f_{5/2, 7/2}$,  $6p_{1/2, 3/2}$ and
$5d_{3/2, 5/2}$ are valence  shells, then the number of valence shells $N_v$ 
and core shells $N_c$ are 7 and 17 respectively. All the orbitals  are made
spectroscopic, that is, each orbital has $n-l-1$ ( where $n$ and $l$  are
the principal and orbital angular momentum quantum numbers) number of 
nodes. The other possibility is calculating the orbitals without imposing
the constraint on the number of nodes. Orbitals of this type are referred to
as  correlation orbitals. These are usually contracted, state specific and 
represent correlation effects of specific states very well. For statistical 
properties calculations, the spectroscopic orbitals are appropriate. The
MCDF  calculations to generate the orbitals include important intra valence
correlation effects  but it is incomplete within the orbital basis considered. 
A configuration interaction (CI)
calculation, which  is diagonalization of $H^{\rm DC}$ matrix, of all the 
possible CSFs can capture
the remaining correlation  effects. However, such a calculation is impossible 
as the size of the CSF space increases
exponentially with  the number of active valence electrons and $N_v$. In our 
calculations
the CSF space consists  of single and double excitations from the ground
configuration
\begin{equation}
   |\gamma_iP_iJ_i\rangle = [(a^{\dagger}_{6p})^{p_i},\nu_{6p}X_{6p}
                            [(a^{\dagger}_{5d})^{q_i},\nu_{5d}X_{5d}
                            [(a^{\dagger}_{6s})^{r_i},\nu_{6s}X_{6s}
                            [(a^{\dagger}_{4f})^{s_i},\nu_{4f}X_{4f}
                            ]^{J_{1i}}]^{J_{2i}}]^{J_{3i}}]^{J_i}
                            |\Phi_0\rangle,
  \label{phii}
\end{equation}
where $p_i + q_i + r_i + s_i = 2 + m + n$,  and impose the conditions
$m\geq s_i \geq m-2$, $2\geq r_i \geq 0$, $n\geq q_i \geq n-2$ and
$2 \geq p_i$. Thus, all the CSFs considered are connected to the ground 
configuration $[{\rm Xe}]6s^24f^m5d^n$. The CSF space has ten
non-relativistic configurations, among these $4f^m5d^2$ and 
$4f^{m-1}6s5d6p$ together contribute more than half of the CSFs. The 
maximum number of CSFs arises from $4f^{m-1}6s5d6p$, which has the largest 
number 
of open shells. The number of CSFs arising from each non-relativistic 
configuration and range of $\langle\gamma PJ| H^{\rm DC}|\gamma PJ\rangle$ are 
given in Table \ref{table1}.

\begin{table}
  \caption{ \label{table1} The number of relativistic CSFs $N_{\rm CSF}$ 
           arising from each of the non-relativistic configurations. The 
           range of the diagonal elements or $\epsilon_k^{\min}$ and 
           $\epsilon_k^{\max}$  are also given, the $\epsilon_k$ values are
           defined relative to the lowest $\epsilon_k$ in the CSF space. In 
           the table, the configurations are defined relative to the ground 
           configuration $4f^m6s^2$, where $m$ is 4, 5 and 6 for Nd, Pm and 
           Sm respectively. 
  }
  \begin{ruledtabular}
  \begin{tabular}{lcccccc}
     & \multicolumn{2}{c}{Nd} 
     & \multicolumn{2}{c}{Pm} 
     & \multicolumn{2}{c}{Sm} \\ \cline{2-7}
    Config & $N_{\rm CSF}$ & $\epsilon_k^{\rm min} - \epsilon_k^{\rm max}$ 
           & $N_{\rm CSF}$ & $\epsilon_k^{\rm min} - \epsilon_k^{\rm max}$ 
           & $N_{\rm CSF}$ & $\epsilon_k^{\rm min} - \epsilon_k^{\rm max}$ \\
     \hline
  $4f^m6s^2$        & 19  & 0.0 -0.25 & 29   & 0.0 -0.28 & 46   & 0.0 -0.38 \\
  $4f^m6s5d$        & 288 & 0.31-0.64 & 521  & 0.09-0.53 & 748  & 0.03-0.58 \\
  $4f^{m-1}6s^26p$  & 40  & 0.39-0.69 & 92   & 0.33-0.65 & 172  & 0.39-0.85 \\
  $4f^m5d^2$        & 608 & 0.63-1.08 & 1090 & 0.24-0.69 & 1579 & 0.18-0.86 \\
  $4f^{m-1}6s5d6p$  & 684 & 0.67-1.04 & 1619 & 0.43-0.87 & 3014 & 0.44-1.00 \\
  $4f^m6p^2$        & 236 & 0.97-1.30 & 409  & 0.83-1.25 & 602  & 0.87-1.38 \\
  $4f^{m+1}6p$      & 172 & 1.11-1.50 & 241  & 1.02-1.40 & 276  & 1.09-1.61 \\
  $4f^{m-2}6s^26p^2$& 31  & 1.14-1.30 & 91   & 1.09-1.42 & 234  & 1.19-1.58 \\
  $4f^{m-2}6s^25d^2$& 78  & 1.38-1.64 & 240  & 1.01-1.34 & 608  & 1.03-1.56 \\
  $4f^{m+2}$        & 46  & 1.56-1.90 & 46   & 1.61-1.97 & 46   & 1.74-2.10  
  \end{tabular}
  \end{ruledtabular}
\end{table}

Consider $N_a$  and $N_{\alpha}$ as the number of active valence electrons 
and single particle states respectively. Then, $N_a$ is 6, 7 and 8 for Nd, 
Pm and Sm respectively and $N_{\alpha} = 32$ since the 
valence space consists of $6s_{1/2}$, $6p_{1/2, 3/2}$, $5d_{3/2, 5/2}$ and 
$4f_{5/2, 7/2}$ ( each orbital contribute $2j+1$ number of single 
particle states ). In addition, define the number of unfilled single particle 
states 
as  $N_o = N_\alpha - N_a$. Then, the number of possible determinantal states 
is ${\tiny \left ( \begin{array}{c} N_\alpha \\ N_a\end{array}\right )}$ and 
the two-electron Coulomb interaction couples each  Slater determinant to 
\begin{equation}
   K = 1 + N_aN_o + N_a(N_a - 1)(N_o)(N_o - 1)/4
 \label{K}
\end{equation}
determinants which include itself and others with shell occupancies different 
by one and two. In a single determinant approximation of the ground state, 
$K$ is also the number of determinants
in the manifold which includes the single and double excitations. This
is the determinantal equivalent of the CSF manifold chosen in our calculations.
For further analysis, consider these $K$ determinants as the manifold chosen
for calculations. 
As mentioned in the context of CSFs, the representation of two-electron 
Coulomb interaction is incomplete in such a determinantal manifold. This is 
because, the Coulomb 
interaction couples pairs of determinants which differ by one or two 
occupancies. Hence, for the singly excited determinants, the zeroth order 
description of Coulomb interaction is incomplete without the triply excited 
ones. The quadruply excited are similarly needed for the doubly excited 
determinants. Among the $K$ determinants, the fraction coupled to a singly 
excited one is
\begin{eqnarray}
    K_s & = &\frac{ K - (N_a-1)(N_a - 2)(N_o - 1)(N_o - 2)/4}{K} \\ \nonumber
        & \sim & \frac {1}{N_a} \;\;\;\;\mbox{when}\;\;\;\; N_\alpha \gg N_a.
  \label{ksng}
\end{eqnarray}
The last term in the numerator is the number of triply excited states 
coupled to a singly excited determinant but absent in the manifold. 
Similarly, the fraction of the determinants coupled to double excited 
determinant is 
\begin{eqnarray}
    K_d & = &\frac{ K - [ (N_a - 2)(N_o - 2) + (N_a-2)(N_a - 3)
                   (N_o - 2)(N_o - 3)/4]}{K} \\ \nonumber
        & \sim & \frac {1}{N_a} \;\;\;\;\mbox{when}\;\;\;\; N_\alpha \gg N_a.
  \label{kdbl}
\end{eqnarray}
That is, each of the states considered are directly coupled to  $\sim 1/N_a$
of the total number. However, higher order two-electron Coulomb interaction 
connects all the determinants 
within the same $J$ manifold. For $N_a\sim 3$, the $K_s$ and $K_d$ scaling  
show the number of
determinants mixed at first order is close to the GOE  predicted value of
number of principal components (NPC). Similar scalings  apply to the CSFs as
these are linear combinations of Slater determinants.  This implies that a
CSF space consisting of single and double excitations  alone can provide a
good representation of the eigenvalues and eigenfunction properties.


\section{Eigenfunction properties in EGOE(1+2)}
\label{egoe}

This section gives a brief introduction to EGOE(1+2) and a summary, for
later use, of the results known for this random matrix ensemble; see
\cite{Ko-01} for a review on this subject and also \cite{Agk-03}.


\subsection{Definition of EGOE(1+2)}

Let us start with two-body embedded GOE, i.e. EGOE(2) or TBRE defined for
spinless fermion systems. The EGOE(2) for $m$ ($m > 2$) fermion with the
particles distributed say in $N$ single particle states $\l| \l. \nu_i
\ran\r.$, $i=1,2,\ldots,N$) is generated  by defining the Hamiltonian H,
which is 2-body, to be GOE in the 2-particle space and then propagating it
to the $m$-particle spaces by using the geometry (direct product structure) of
the $m$-particle  spaces. Operator form for a 2-body Hamiltonian H=V(2) is
defined by,
\begin{eqnarray}
  && V(2) =  \dis\sum_{\nu_i < \nu_j,\;\nu_k < \nu_l} \lan \nu_k\;\nu_l
     \mid V(2) \mid \nu_i\;\nu_j\ran_a\;a^\dg_{\nu_l}\,a^\dg_{\nu_k}\,
     a_{\nu_i}\,a_{\nu_j}\;;  
  \label{v2sym}  \\
  && \lan \nu_k\;\nu_l \mid V(2) \mid \nu_j\;\nu_i\ran_a = - \lan \nu_k\;\nu_l
     \mid V(2) \mid \nu_i\;\nu_j\ran_a\;, \nonumber \\
  && \lan \nu_k\;\nu_l \mid V(2) \mid \nu_i\;\nu_j\ran_a =  \lan \nu_i\;\nu_j
     \mid V(2) \mid \nu_k\;\nu_l\ran_a   \nonumber
\end{eqnarray}
Now, the Hamiltonian H matrix in m-particle spaces, in the occupation number
basis (occupation numbers will be $0$ or $1$ and the basis is generated by
distributing the $m$ particles in all possible ways in the single particle
states $\nu_r$), is defined in terms of the  two-body matrix elements $\lan
\nu_k\;\nu_l \mid V(2) \mid \nu_i\;\nu_j\ran_a$ (note that the subscript `a'
stands for antisymmetrized two-particle states) and the non-zero matrix
elements are of  three types,
\begin{eqnarray}
  && \lan \nu_1 \nu_2 \cdots \nu_m \mid V(2) \mid \nu_1 \nu_2 \cdots
     \nu_m \ran_a = \dis\sum_{\nu_i < \nu_j \leq \nu_m}\; \lan \nu_i
     \nu_j \mid V(2) \mid \nu_i \nu_j \ran_a    \nonumber \\
  && \lan \nu_p \nu_2 \nu_3 \cdots \nu_m \mid V(2) \mid \nu_1 \nu_2 \cdots
     \nu_m \ran_a = \dis\sum_{\nu_i= \nu_2}^{\nu_m}\; \lan \nu_p
     \nu_i \mid V(2) \mid \nu_1 \nu_i \ran_a    \nonumber \\
  && \lan \nu_p \nu_q \nu_3 \cdots \nu_m \mid V(2) \mid \nu_1 \nu_2
     \nu_3 \cdots \nu_m \ran_a = \lan \nu_p
     \nu_q \mid V(2) \mid \nu_1 \nu_2 \ran_a \;\;;  
  \label{v2ele}   \\
  && \mbox{all other}\;\;\lan \cdots \mid H \mid \cdots \ran_a = 0\;\;
  \mbox{due to the two-body selection rules}.   \nonumber 
\end{eqnarray}
The EGOE(2) is defined by the above Eqs. and a GOE representation for V(2),
\begin{eqnarray}
   && \lan \nu_k\;\nu_l \mid V(2) \mid \nu_i\;\nu_j\ran_a\;\; 
      \mbox{are independent Gaussian random variables}  \nonumber \\
   && \overline{\lan \nu_k\;\nu_l \mid V(2) \mid \nu_i\;\nu_j\ran_a} = 
      0\;,\;\;\;   \nonumber \\
   && \overline{\l|\lan \nu_k\;\nu_l \mid V(2) \mid \nu_i\;\nu_j\ran_a\r|^2}=
      v^2(1+\delta_{(ij),(kl)}) 
   \label{v2var}
\end{eqnarray}
where the bar denotes ensemble average and $v^2$ is a constant. Note that  
${\tiny{d(m)=\l(\barr{c} N \\ m\earr\r)}}$ is the Hamiltonian matrix 
dimension and the number of independent two-body matrix elements is 
$[d(2)(d(2)+1)]/2$.

Hamiltonian for realistic systems such as  atoms consists of a mean-field
one-body (defined by a finite set of single particle  states) plus a
complexity generating two-body interaction. Then the appropriate random
matrix ensemble, for many purposes, is EGOE(1+2) defined by  
\begin{equation}
  \{H\} = h(1) + \lambda \{V(2)\} 
  \label{hensbl}
\end{equation}
where $\{\;\}$ denotes an ensemble. In (\ref{hensbl}), the mean-field one-body
Hamiltonian $h(1)=\sum_i \epsilon_i n_i$ is a fixed one-body operator
defined by the single particle energies $\epsilon_i$ with average spacing
$\Delta$ (note that $n_i$ is the number operator for the single particle
state $\l. \l| i\r. \ran$). In general one can choose $\epsilon_i$ to form 
an ensemble; see
\cite{Ja-01} for examples.  The $\{V(2)\}$ in (\ref{hensbl}) is EGOE(2) with 
$v^2=1$ in Eq. (\ref{v2var}) and $\lambda$ is the strength of the two-body 
interaction (in units of $\Delta$). Thus, EGOE(1+2) is defined by the four 
parameters $(m,N,\Delta,\lambda)$ and without loss of generality we choose 
$\Delta=1$. Construction of EGOE(1+2) follows from Eqs. (\ref{v2sym}), 
(\ref{v2ele}) and (\ref{v2var}) by just adding to the diagonal matrix elements 
(see the first equality in (\ref{v2ele})) the term $\sum_i\,
\epsilon_{\nu_i}$ where $\nu_i$ are the occupied single particle states for
the given basis state.


\subsection{Basic properties of EGOE(1+2)}

Most significant aspect of EGOE(1+2) is that as $\lambda$ changes, in terms of
state density, level fluctuations, strength functions and entropy, the ensemble
admits three  chaos markers as described in Fig. 1 and reviewed in
\cite{Agk-03}. Firstly, it is well known, via the EGOE(2) results in
\cite{Mo-75,Br-81} and the fact that general $h(1)$'s produce Gaussian
densities, that the state densities $\rho^{H,m}(E)=\lan \delta(H-E)\ran^m$ take
Gaussian form, for large enough $m$, for all $\lambda$ values (often the
superscripts $(H,m)$ are dropped), 
\begin{subequations}
   \begin{equation}
      \rho^{H,m}(E) = \dis\frac{1}{\dis\sqrt{2\pi}\,\sigma_H(m)}
      \,\exp-\frac{{\hat{E}}^2}{2},
   \end{equation}
   \begin{equation}
      \hat{E}=(E-\epsilon_H(m))/ \sigma_H(m)
   \end{equation}
   \label{rhoh}
\end{subequations}
In (\ref{rhoh}), $\epsilon_H(m)=\lan H \ran^m$ is the spectrum centroid and 
similarly $\sigma_H(m)$ is the spectrum width. In practice there will be 
deviations from the Gaussian form and they are taken into account by using 
Edgworth expansion \cite{Ke-69} in terms of the skewness ($\gamma_1$) and 
excess ($\gamma_2$) parameters. With 
$\rho^H(E)dE = \eta(\widehat E)d\widehat{E}$,
the $\eta_{\cal G}(\widehat E)$ with lower order Edgeworth corrections is, 
\begin{equation}
   \begin{array}{l}
     \displaystyle
     \eta_{\rm ED}(\widehat{E}) = \\
     \displaystyle
     \eta_{\cal G}(\widehat{E})\left \{ 1 + 
     \frac{\gamma_1}{6}He_3(\widehat{E}) +
     \frac{\gamma_2}{24}He_4(\widehat{E})     +
     \frac{\gamma_1 ^2}{72}He_6(\widehat{E}) \right \}
   \end{array}
 \label{edge}
\end{equation}
where $He_r(\widehat{E})$ are Hermite polynomials. With $\lambda$
increasing, there is a chaos marker $\lambda_c$ such that for $\lambda \ge
\lambda_c$ the level fluctuations follow GOE, i.e. $\lambda_c$ marks the
transition in the nearest neighbor spacing distribution from Poisson to
Wigner form. Parametric dependence of $\lambda_c$ is $\lambda_c \propto
1/m^2N$ \cite{Ja-97}. Now let us define strength functions.

To maintain consistency with the notations used in EGOE(1+2) literature,
from hereafter, we represent CSFs as $|k\rangle$ and ASFs as $|E\rangle$.
Then, Eq. \ref{asf} assumes the form $|E\rangle = \sum_k c_k^E|k\rangle$.
Given the mean-field $h(1)$  basis states $\l.\l| k \ran \r.=\sum_E\,C^E_k
\l.\l| E \ran \r.$,  the strength functions (one for each $k$) $F_k(E)
=\sum_{\beta \in E}\,   \l|C^{E,\beta}_k\r|^2 = \overline{\l|C^E_k\r|^2}\,(d
\rho^H(E))$. As $\lambda$ increases further from $\lambda_c$, the strength
functions change from Breit-Wigner (BW) to Gaussian form and the transition
point is dented by $\lambda_F$; see Fig. 1. The BW and Gaussian (denoted by 
$\cg$) forms of $F_k(E)$ are,
\begin{subequations}
  \begin{equation}
    F_{k:BW}(E) = \dis\frac{1}{2 \pi}\, \dis\frac{\Gamma_k}{(E-E_k)^2 +
    \Gamma_k^2/4},
  \end{equation}
  \begin{equation}
  F_{k:\cg}(E) = \dis\frac{1}{\dis\sqrt{2
    \pi}\,\sigma_k} \,\exp-\frac{(E-E_k)^2}{2 \sigma_k^2}
  \end{equation}
  \label{fkgss}
\end{subequations}
where $E_k = \lan k | H | k \ran$. With $p=\int^{{\ce}_p}_{-\infty}\,F_k(E)
dE$, the spreading width $\Gamma_k={\ce}_{3/4} - {\ce}_{1/4}$.  Similarly
the variance of $F_k$ is $\sigma_k^2=\lan k \mid H^2 \mid k \ran - (\lan k
\mid H \mid k  \ran)^2$. For $\lambda_c \leq \lambda \leq \lambda_F$ (this
is called BW domain) they are close to BW form and for $\lambda >
\lambda_F$  (this is called Gaussian domain)  they approach Gaussian form.
In fact the BW form starts in a region  below $\lambda_c$  (as shown in Fig.
1, there is a $\lambda_0$ such that below $\lambda_0$, the strength
functions are close to $\delta$-function form and for $\lambda > \lambda_0$
there is onset of BW form) but the fluctuations are close to Poisson for
$\lambda < \lambda_c$. Arguments based on BW spreading widths give
$\lambda_F \propto 1/\sqrt{m}$ \cite{Fl-97,Ja-02} and this result will be
used later. Unlike the Poisson to GOE transition in level fluctuations,
which is common for Hamiltonians with regular and irregular (chaos
generating) parts, the BW to Gaussian transition in strength functions is a
unique signature for the  operation of EGOE(1+2) in finite quantum systems.
For the BW to Gaussian transition, a good interpolating  function has been
constructed recently \cite{Agk-03}, 
\begin{equation} 
  \barr{l}
    F_{k:BW-\cg}(E:\alpha,\beta)\,dE = \\
    \dis\frac{(\alpha\beta)^{\alpha-\frac{1}{2}}\;\Gamma(\alpha)}{\dis\sqrt{
    \pi}\;\Gamma(\alpha-\frac{1}{2})} \;\dis\frac{dE}{\l((E-E_k)^2+\alpha
    \beta\r)^\alpha}\;,\;\;\;\alpha \ge 1 
  \earr 
  \label{fitfun} 
\end{equation}
In $F_{k:BW-\cg}(E: \alpha,\beta)$, $\beta$  supplies the scale while
$\alpha$ defines the shape (hence $\alpha$ is the significant parameter) as
it gives BW for $\alpha=1$ and Gaussian for $\alpha \rightarrow \infty$. 
The variance of $F_{k:BW-\cg}$ is  $\sigma^2(F_{k:BW-\cg}) = \sigma_k^2 = 
\alpha \beta
/(2 \alpha -3)$,  $\alpha > 3/2$. It is important to stress that in general
$(\alpha,\beta)$ change with $k$ although for EGOE(1+2) they are nearly
constant except for $F_k$ with $E_k$ very far from their centroid. Fig. 2
shows that the interpolating function (23) fits very well the numerical 
EGOE(1+2) results for strength functions; for more detailed discussion see
\cite{Agk-03}.

One important measure of the complexity of eigenstates of  interacting
systems is the participation ratio PR (denoted by $\xi_2$),  $\xi_2(E)  = 
\{\sum_k\,|C_k^E|^4\}^{-1}$ where the subscript `2' for $\xi$ denotes that
it is the second R\`{e}nyi  entropy \cite{Im-02}; in literature PR is also
often called number of  principal components (NPC) as it gives the number of
basis states that make up the eigenstate with energy $E$. Following Ref.
\cite{Ks-01} one can write $\xi_2$ in terms of the strength functions,
\begin{equation}
   \l\{\xi_2(E)/\xi_2^{GOE}\r \}^{-1}  = \dis\frac{1}{\l[\rho^H(E)\r ]^2}\;
   \dis\int_{-\infty}^{\infty}\,dE_k\;\rho^{\bf{h}}(E_k) \l[F_k(E)\r ]^2\,,
   \;\;\; \xi_2^{GOE}  =   d/3
  \label{xi2goe}
\end{equation}
In writing (\ref{xi2goe}), it is assumed that the level and strength 
fluctuations are of GOE type and hence it is valid only for 
$\lambda > \lambda_c$. Further, $\rho^{\bf{h}}(E_k)$ is the density of $E_k$ 
energies and it is generated by an effective one-body Hamiltonian ${\bf h}$ 
\cite{Ks-01}. Also, in general $\rho^{\bf{h}}(E_k)$ is also a Gaussian for 
EGOE(1+2). Assuming that $\alpha$ and $\beta$ are $k$-independent, $\xi_2(E)$ 
in the BW to Gaussian transition domain can be evaluated by substituting 
$F_{k:BW-\cg}$ for $F_k(E)$ in (\ref{xi2goe}). At the spectrum center the 
integral in (\ref{xi2goe}) can be evaluated and it gives \cite{Agk-03},
\begin{equation}
  \barr{l}
    \xi_2(E=0)/\xi_2^{GOE} =  
    \l\{ \dis\sqrt{\dis\frac{2}{(2 \alpha -3)}}\;
    \dis\frac{\Gamma^2(\alpha)}{\Gamma^2(\alpha-\frac{1}{2})}\;
    \dis\frac{1}{\dis\sqrt{\zeta^2 (1-\zeta^2)}} \; \times \r .\\
    \l . U\l(\frac{1}{2},\; 
    \frac{3}{2} - 2 \alpha,\; \frac{(2\alpha-3)(1-\zeta^2)}{2 \zeta^2}\r)
    \r\}^{-1}
  \earr
  \label{xi2zero}
\end{equation}
where $U(---)$ is hypergeometric-U function \cite{Ab-64}. In (\ref{xi2zero}), 
the correlation coefficient  $\zeta = \sqrt{ 1 -
\overline{\sigma_k^2}/\sigma_H^2 }$ where $\overline{\sigma_k^2} =
(1/d)\sum_{i\neq j} H_{ij}^2$ and $\sigma_H^2 = (1/d)\sum_i (E_i -
\epsilon)^2$ are the variance of the off-diagonal Hamiltonian matrix
elements and eigenvalues respectively. However, in the Gaussian domain, with
$F_k(E)$ being a Gaussian, gives for any $E$,
\begin{equation}
  \xi_2(E)/\xi_2^{GOE} = \dis\sqrt{1-\zeta^4}\;
  \exp-\frac{\zeta^2 {\hat{E}}^2}{1+\zeta^2}
  \label{goexi2}
\end{equation}
Instead of NPC, it is possible to use the closely related information
entropy $S^{info}(E)$; $S^{info}(E)=-\sum_E \,\l|C_k^E\r|^2 \log
\l|C_k^E\r|^2$. An interesting recent observation is that \cite{Ks-02}  as
we increase $\lambda$ much beyond $\lambda_F$, there is a chaos marker
$\lambda_t$ around which different definitions of entropy (for example
$S^{info}$, thermodynamic entropy defined via $\rho^H(E)$, single particle
entropy defined via occupation numbers),  temperature etc. will coincide and
also strength functions in $h(1)$ and $V(2)$ basis will coincide. Thus
$\lambda \sim \lambda_t$ region is called the thermodynamic region and
\cite{Ko-03} gives first application of this marker. 

It is important to point out that the eigenstates of atoms carry good
angular momentum ($J$) (in some situations even good $L,S$) and therefore 
in principle one should consider EGOE(1+2) preserving $J$-symmetry (called
EGOE(1+2)-$J$ in  \cite{Ko-01}). Theory for EGOE(1+2)-$J$ is not available
(see however \cite{Kk-02,Pw-04} for some first attempts), and just as in
the  nuclear shell model studies \cite{Ko-01,Go-01}, it is assumed that the
forms for $\rho^H(E)$, $F_k(E)$, (NPC)$_E$ etc. derived using EGOE(1+2) for
spinless fermion systems, extend to $J$ spaces. This can be considered to
be an aspect of chaos in atoms and this is the basis for statistical atomic
spectroscopy as being developed by Flambaum; see \cite{Fl-99,Ks-02a,Fl-02}
and references therein.  


\section{Results for $\rho^H(E)$, $F_k(E)$, $\xi_2(E)$ and $\lan n_\alpha 
\ran^E$}
\label{results}

In the present study, we select the $J= 4^+$ CSF manifold for Nd and Sm 
atoms. The parity is chosen equal to the parities of the ground
configurations $4f^46s^2$ and $4f^66s^2$ respectively and the total
angular momentum $J$ is  the same as in the Ce I work \cite{Fl-94}. For
Pm, which has an odd number (i.e. 7) of  active valence electrons we choose
$J=9/2^-$ CSFs. The parity is again  equal to the ground configuration
parity $4f^56s^2$. It is to be reiterated  that, the orbitals of the
present study are different from our earlier work  \cite{Dk-03}. As
mentioned in Section \ref{orb-csf}, the orbitals in this work are
calculated in a multiconfiguration potential, whereas in \cite{Dk-03},
the orbitals are calculated in a single configuration potential. The
manifolds of the Nd, Pm and Sm have 2200, 4375 and 7325 CSFs
respectively. The CSFs are  generated in the sequence of sub-shell
occupations but later, these are energy  ordered. The CI calculation is
carried out within the whole space for Nd and  Pm. For Sm, as mentioned
in \cite{Dk-03}, the density of states $\rho(E)$ is  bimodal. We select
the first 6300 CSFs which contribute to the first peak for  the CI
calculation. The atomic state  $|E\rangle $ of the Nd, Pm, and Sm 
obtained from the CI calculations have energy range of 2.20, 2.27 and
1.60  hartree respectively.


\subsection{Density of states $\rho^H(E)$}
\label{density}

By definition, density of state $\rho(E)= \langle \delta(H-E)\rangle $ is a 
sequence of 
$\delta$-functions. For obtaining the smoothed (with respect to energy $E$)
density of states and  for comparison with the EGOE(1+2) predictions of
$\rho^{H, m}(E)$, we  construct  the binned density of states. In the our
calculations, the energy range of $|E\rangle$ of each atom is divided into
sixty bins of equal sizes.  The density of states is then binned and
normalized, these are shown in  Fig.\ref{ndpmsm_den}.  The Gaussian and
Edgeworth  corrected Gaussian calculated from the centroid $\epsilon$, 
variance $\sigma^2$, skewness $\gamma_1$ and  excess $\gamma_2$ of $\rho(E)$
are also shown in the figure ( $\epsilon$,  $\sigma^2$, $\gamma_1$ and
$\gamma_2$ values are given in the figure).  For SmI Gaussian is in good 
agreement
with the $\rho(E)$ except at and around the  centroid. At the centroid the
Gaussian is $\sim$ 13\% less than the actual $\rho(E)$. In general all
$\rho(E)$ are multimodal in  structure and it is most prominent in Nd. The
$\rho(E)$ for Nd has a pronounced  peak at $\hat{E} \sim -1.5$ and 
fluctuations above the  mean for $\hat{E}$ between $1.0$ and  $2.5$.
Similarly for Pm, $\rho(E)$ has a pronounced peak at $\hat{E} \sim 1.4$. As 
discussed ahead, the multimodal structures correlate with similar 
structures of the number of principal components and observed deviations of 
the strength function from the $\epsilon_k$ dependent variation. Finally for
SmI, there is a peak at $\hat{E} \sim 2$.

The multimodal form of $\rho^{H, m}(E)$ can be understood by decomposing it
into a sum of  partial densities. For each sub-shell occupancy
$\widetilde{m}$ one can define a partial density $\rho^{H,
\widetilde{m}}(E)$ with centroids $E_c(\widetilde{m})$ and $\sigma^2(
\widetilde{m})$ defined by,
\begin{equation}
   E_c(\widetilde{m}) = \lan H \ran ^{\widetilde{m}}\;,\;\;\; 
   \sigma^2(\widetilde{m})= \lan (H-E_c(\widetilde{m}))^2\ran^{\widetilde{m}}\;.
\end{equation} 
Note that $\sigma^2(\widetilde{m})$ is sum of partial variances $\sigma^2(
\widetilde{m} \rightarrow \widetilde{m}^\prime)$ where
\begin{equation}
  \barr{c}
    \sigma^2(\widetilde{m}) = \dis\sum_{\widetilde{m}^\prime}\;
    \sigma^2(\widetilde{m} \rightarrow \widetilde{m}^\prime)\;, \\ \\
    \sigma^2(\widetilde{m} \rightarrow \widetilde{m}^\prime) =
    \l\{d(\widetilde{m})\r\}^{-1}\;\dis\sum_{\alpha,\beta}\;\l|\lan 
    \widetilde{m}\; \alpha \mid H \mid
    \widetilde{m}^\prime\;\beta\ran\r|^2\;\;\mbox{for}\;\; \widetilde{m} \neq 
    \widetilde{m}^\prime \\
    \sigma^2(\widetilde{m} \rightarrow \widetilde{m}) =
    \l\{d(\widetilde{m})^{-1}\;\dis\sum_{\alpha,\beta}\;\l|\lan 
    \widetilde{m}\; \alpha \mid H \mid
    \widetilde{m}\;\beta\ran\r|^2\r\} - \l\{E_c(\widetilde{m})\r\}^2
  \earr
\end{equation}
Usually the $\sigma^2(\widetilde{m} \rightarrow \widetilde{m})$ is called
internal variance and others, partial variances \cite{Ch-71}. In terms of
the partial  densities the total density of states is,
\begin{equation}
   \rho^{H, m}(E) = [d(m)]^{-1}\;\sum_{\widetilde{m}} 
                    \rho^{H, \widetilde{m}}(E)\; d(\widetilde{m})
  \label{partial}
\end{equation}
where $m$ is the number of active valence electrons, $\widetilde{m}$ 
represent the  possible sub-shell occupancies, $d(m)$  and $d(\widetilde{m})$
are the  number of CSFs in  the CI calculation and number of CSFs in the
$\widetilde{m}$ sub-shell occupancy  manifold respectively. That is,
$\rho^{H, m}(E)$ is the sum of partial  densities $\rho^{H,
\widetilde{m}}(E)$ from the sub-shell occupations defined  in Eq.
(\ref{phii}). It is important to stress that Eq. (\ref{partial}) is exact. 
As evident from Eq. (\ref{phii}), each of the possible sub-shell occupations 
have different angular momentum coupling sequence. It is easily seen that
$\rho^{H,m}(E)$ takes multimodal shape when the partial  densities are well
separated with weak mixing between them (i.e. partial variance is much
smaller than the internal variance). The range of $\epsilon_k$ (i.e. 
diagonal elements of  the Hamiltonian matrix $H_{rr}^{\rm DC}$ ) of each
non-relativistic  configuration given in Table \ref{table1}, which is a
representation of the  partial density spread, is plotted for Nd in  Fig.4a.
At  $\sim 0.62$ hartree, it shows small overlap between the
$4f^46s5d$ and next in energy $4f^45d^2$ and  $4f^36s5d6p$  configurations.
The separation is also visible in the Hamiltonian  structure discussed ahead
[Fig. \ref{fighmat} ahead shows two distinct blocks  centered around $\sim
0.5$ hartree and $\sim 0.8$ hartree respectively]. In addition, the  Coulomb
mixing $\langle 4f^46s5d|1/r_{12}|4f^45d^2\rangle$ and  $\langle
4f^46s5d|1/r_{12}|4f^36s5d6p\rangle$ are weak as the configurations differ by
a single occupancy. The separation between the configurations  persists in 
$|E\rangle$ and the contributions from $4f^46s5d$ configuration manifests as 
the smaller peak of the bimodal $\rho^{H, m}(E)$. The energy range of the
remaining configurations show large overlaps and  contribute to the main peak
of the Nd $\rho^{H, m}(E)$. The  $4f^36s^26p$  configuration which lie
between $4f^46s5d$ and $4f^45d^2$ energetically,  is single replacement from
$4f^45d^2$ and doesn't contribute to smoothing the $\rho^{H, m}(E)$. 
Fig.4b  shows the result of Eq. (\ref{partial}) with
partial  densities represented by Gaussians for Nd. The agreement between
theory and calculations is excellent. This demonstrates that it is possible
to construct the total density of states as sums of Gaussian partial
densities (even though a single Gaussian representation as shown in Fig.
\ref{ndpmsm_den} deviates strongly from the calculated results). This result
is significant as it is possible to calculate the centroids and variances
(also partial variances) without recourse to $H$ matrix construction
\cite{Ch-71}. Similar to the Nd, the bimodal structure of the Pm's $\rho^{H,
m}(E)$ is a  consequence of the small overlap between the  partial densities
of the $4f^46s5d6p$ and $4f^56p^2$ configurations. However, unlike Nd, these 
configurations are in the higher end of the energy range. In contrast to  Nd
and Pm, the Sm atom has partial densities with large overlap and  the
$\rho^{H, m}(E)$ is without prominent secondary peaks.  Now we will turn to
strength functions which are nothing but the partial densities with 
$\widetilde{m}$'s divided into individual basis states.


\subsection{Strength functions $F_k(E)$}

  The $F_k(E)$ of the individual CSFs $|k\rangle$ exhibit large  fluctuations
and significant variation is observed between $F_k(E)$ of  neighboring
$|k\rangle$. This is evident from the selected $|k\rangle$  presented in our
Sm I work \cite{Dk-03}. For statistical description we consider the 
representative strength function $\overline{F_k(E)}$ which is the average of
individual $F_k(E)$ within the range of energy  $\epsilon_k \pm \Delta
\epsilon$. In the present work, we have chosen $\widehat{\epsilon}_k \pm
.25\sigma_k$ as the range of averaging to calculate $\overline{F_k(E)}$  and
the centroids of the individual  $F_k(E)$ are aligned while averaging. Fig.
\ref{ndpmsm_fk} shows $\overline{F_k(E)}$ of Nd I, Pm I and SmI calculated at
$\widehat{\epsilon}_k = -1.5,  -1.0, -0.5,$ and $0$. For all the three
atoms,  the interpolating function of Eq. (\ref{fitfun}) fits the
$\overline{F_k(E)}$ very well; the $\beta$ parameter in Eq. (\ref{fitfun}) has 
been eliminated using the calculated $\sigma_k$. The best fit value of $\alpha$ 
for Nd I, Pm I and SmI for $\widehat{\epsilon}_k = 0$ are 5, 7.5 and 11
respectively, thus these are in the Breit-Wigner to Gaussian transition
region with SmI closest to Gaussian. This result is already reported in
\cite{Agk-03} (it should be noted that in this paper $\overline{F_k(E)}$ is
constructed by averaging $F_k(E)$ over 3\% of the basis states around
$\widehat{\epsilon}_k = 0$ and this is different from the procedure used in
Fig. 5 and therefore there is slight difference in the $\alpha$ values
extracted). Moving away from $\widehat{\epsilon}_k = 0$, firstly it is seen
that the parameter $\alpha$ in Eq. (\ref{fitfun}) is $k$ dependent and it
changes from a value close to that of BW to a value giving close to Gaussian
form as  $\widehat{\epsilon}_k$ is increasing from -1.5. For Nd, Pm and Sm 
$\alpha$ changes from 2 to 5.2,  2.5 to 7.5 and 2.5 to 10.6 respectively. 
The variation in $\alpha$ seen in Fig. 5 is understood from the fact that
near the Fermi surface (i.e. for  $\widehat{\epsilon}_k \sim -1.5$ to $-1$)
the levels are well separated and hence the mixing is weak giving BW form and
as we go towards the center the mixing is strong giving close to Gaussian (or
between BW and Gaussian) form. It should be pointed out that EGOE(1+2)
ensemble do not produce large changes in $\alpha$ as $k$ changes in $F_k(E)$. However, in all the
calculations it is seen that $\sigma_k$ is essentially constant as expected
from EGOE(1+2). Clearly, the $F_k(E)$ analysis shows that modifications of
EGOE(1+2) are needed.


\subsection{Number of principal components $\xi_2(E)$  and localized states}
\label{npc}

The inverse participation ratio $\xi_2(E)$ of an eigenstate is the  effective
number of basis functions contributing to it. It provides a  measure for the
presence of  chaos in the system. For a GOE with dimension $d$, the NPC is
$d/3$ independent of energy while for EGOE(1+2) there is strong energy
dependence (it is Gaussian in the Gaussian domain; see Eq. (\ref{goexi2})). The
$\xi_2(E)$ of Nd, Pm and Sm for all $|E\rangle$ are calculated and shown  in
Fig. \ref{npcfig}a-c. The $\xi_2(E)$ of Nd and Pm are multimodal and
localized states (i.e. states with NPC much smaller than GOE or EGOE(1+2)
predicted values) are prominent in all the atoms. We should add that
localized states are also observed  in the previous studies on Ce I
\cite{Fl-94} and Pr I \cite{Cu-01}. In addition, more recently, localization 
quite similar to those shown Fig.6 are also seen in $U^{28+}$ calculations
\cite{Gr-04}. Despite localization, Nd and Pm have 
Breit-Wigner like envelopes, whereas for Sm it is like Gaussian. Nd and Pm 
have Gaussian  like secondary peaks
around $\sim -1.5$ and $\sim 1.4$ respectively and it is to  be noted that,
these are the locations where $\rho^{H, m}(E)$ has local peaks. For Sm, like
$\rho^{H, m}(E)$, it is unimodal.  At the
centroid of the envelopes, the value of $\xi_2(E)$ are $\sim 0.39$,  $\sim
0.42$ and $\sim 0.5$ for Nd, Pm and Sm respectively. These should be compared
with the EGOE(1+2) values given by Eq. (\ref{xi2zero}) (for the Nd, Pm and Sm 
atoms, $\zeta^2$ is 0.88, 0.86 ans 0.83 respectively) and they are 0.44, 0.48 
and 0.55 respectively. Here it is assumed that $\alpha$ value at the 
$\hat \epsilon _k = 0$ can be used for all $\hat\epsilon _k$'s. 
However this is not a good approximation as seen from Fig.6. In a better 
calculation we use Eq. (\ref{xi2goe}), where the integral is divided into seven
segments each with $\hat\epsilon _k$ spread .5 in the range -1.75 to 1.75. 
Within each segment $F_k(E)$ is constructed with $\alpha$ taken for the mid 
$\hat\epsilon _k$ but incorporating exact $F_k(E)$ centroids 
and widths. This calculation gives overall good description for Sm while for 
Pm there are deviations for $\hat\epsilon _k > 0$. This is not 
surprising because there is a secondary peak for $\hat E > 0$, which
cannot be accounted in the EGOE(1+2) model adopted. For Nd we have not shown 
the result as the $\alpha$ for $\hat\epsilon _k = -1.0$ is  
undetermined. Assuming an interpolating value of $\alpha$, it is seen that the
calculations describe reasonably well $\xi _2(E)$ for $\hat E < 0$. 

 In Fig. \ref{npcfig}d-f  $\xi_2(E)$ is binned to remove local
fluctuations and the resulting histograms are shown.
They give at the center the $\xi_2(E)$ to be 0.27, 0.33 and 0.48 for Nd, Pm
and Sm respectively.  Finally, though the number
of  localized states appear numerous in the plots, they are in fact not that
many. This is seen clearly from Fig. \ref{npcfig}g-i where the relative
density of localized states is plotted as a histogram. We classify a state as
localized if it has $\xi_2(E)$ less than or equal to $20\%$ of the binned 
value in corresponding energy range. As shown in Fig. \ref{npcfig}g-i the 
number of localized states in each energy bin $\rho_{\rm loc}(E)$ is 
small for all the atoms studied. This indicates the difference between 
the envelope and binned $\xi_2(E)$ is due to high frequency fluctuations and
not due to purely localized  states. In order to understand more about
localized states an analysis is carried out in terms of occupancies of single
electron orbits and we will turn to this now.


\subsection{Occupation number $\langle n_\alpha \rangle ^E$ and 
            the localized states}

  In the statistical description of a many-body quantum system, properties
of the system are often expressed as functions of the sub-shell occupation numbers
\cite{Fl-02}.  In this section, we use sub-shell occupation numbers to study 
the nature of the localized states.  By definition the occupation number of the 
sub-shell $\alpha$ in $|k\rangle$ is
\begin{equation}
  \langle n_\alpha\rangle _k = \langle k| a^\dagger _\alpha a_\alpha|k\rangle 
\end{equation}
where $a^\dagger _\alpha$ and $a_\alpha$ are the particle creation and 
annihilation operators in the sub-shell $\alpha$ and these are defined in 
Section \ref{mcdf}. Further, we can also define the sub-shell occupancy of 
an atomic state 
\begin{equation}
  \langle n_\alpha\rangle^E = \langle E| a^\dagger _\alpha a_\alpha|E\rangle .
                            = \sum_k |C_k^E|^2 \langle n_\alpha\rangle _k .
\end{equation}
It is evident from  $\xi_2(E)$ calculations 
that, there are several localized states as shown in Fig. \ref{npcfig}g-i. 
Each of these have $\xi_{\rm loc}(E)\ll \overline {\xi_2(E)}$ and few 
$|k\rangle$ can represent 
these states. A possible origin of localized states is strong Coulomb
mixing between few $|k\rangle$ of similar sub-shell occupancies, then
the two-electron Coulomb integral is between same orbitals and is large 
( $K=0$ multipole is allowed). This is indeed the case and the signature is 
the anticorrelation in the trend of $\xi_2(E)$ and occupation number of 
selected sub-shells shown in Fig. \ref{occfig}. The figures show $\xi_2(E)$
and $\langle n_\alpha\rangle ^E$ of few $|E\rangle$ around centroid energy. 
In Fig. \ref{occfig}a, there is anticorrelation between the 
$\xi_2(E)$ and  
$\langle n_{5d_{3/2}}\rangle ^E + \langle n_{5d_{3/2}}\rangle ^E $ of Nd, 
similar trend is observed for Pm in Fig. \ref{occfig}b {\em i.e.} the 
occupancies are large when NPC is small. 
This shows the localized states of Nd and Pm within the spectral 
range considered arise from strong mixing between $|k\rangle$ of high 
$5d$ occupancy. Whereas, the localized states of Sm 
as shown in Fig. \ref{occfig}c arise from the strong mixing between 
$|k\rangle$ of high $6s$ or $6p$ occupancies. 

  To compare and contrast the chaotic and localized states, $|C_k^E|^2$ of 
pair of neighboring chaotic and localized states of Nd, Pm and Sm are shown in 
Fig. \ref{egfn}. The figures clearly shows that the localized states 
( energy states, $\Gamma = 800$, $\Gamma = 2148$ and $\Gamma = 2885$ of Nd, Pm 
and Sm 
respectively) are represented by few $|k\rangle$, whereas the neighboring 
chaotic states ( $\Gamma = 801$, $\Gamma = 2149$ and $\Gamma = 2886$ of Nd, 
Pm and Sm respectively) has contributions from several $|k\rangle$. The
$|C_k^E|^2$ of the chaotic states exhibit a systematic growth and decay, 
in contrast localized states show no systematic trends. To show the 
anticorrelation between $\langle n_\alpha\rangle$ of selected sub-shells and
$\xi_2(E)$, the contributions from each basis 
$|C_k^E|^2\langle n_\alpha \rangle _k$ are plotted. For the localized states
the plots are for $\alpha$ that has the highest value. For the chaotic
states, all the sub-shells show similar trend and approximately an order of 
magnitude less ( see the scales in Fig. \ref{egfn}) than that of the localized 
states.


\section{Hamiltonian matrix structure}
\label{ham_str} 

Understanding the structure of  the Hamiltonian matrix is crucial for
developing an appropriate random matrix model for the Lanthanide atoms. The
diagonal  Hamiltonian matrix elements $H^{\rm DC}_{kk}$ will have
contributions from both the one and two particle Hamiltonian terms. However,
the off diagonal terms $H^{\rm DC}_{kk^\prime}$ have non-zero contributions
from the  Hamiltonian terms  depending on the relative excitation  between
$|k\rangle$ and  $|k^\prime\rangle$. If the two CSFs are singly excited with
respect to each other, then one and two particle terms contributes, whereas
only the  two particle term contributes when the two CSFs are doubly excited
with respect to each other. The off diagonal matrix elements are zero for
triple  or higher excitations. As a consequence, the Hamiltonian matrix is
sparse. In  our present calculations, several of the CSFs which are doubly
excited with respect to ground configuration are quadruply excited with
respect to  each other. For example the configurations
$4f^{m-2}5d^{n+2}6s^2$  and $4f^{m}5d^{n}6p^2$ are doubly excited with
respect to the ground  configuration $4f^m5d^n6s^2$, however they are
quadruply excited with respect  to each other.  Hence, all the matrix
elements between CSFs arising from  $4f^{m-2}5d^{n+2}6s^2$ and
$4f^{m}5d^{n}6p^2$ are zero. There are other configurations similar to these
in the CI space considered. Due to these configurations, the Hamiltonian
matrix in our calculations is sparse,  though we consider only the single
and double excitations from the ground  state configuration. For bringing
out a coarse grained structure of the  Hamiltonian matrix, the square of
matrix elements   $(H^{\rm DC}_{kk^\prime})^2$ (to remove the phase factor)
is calculated and binned in terms of the diagonal elements $H^{\rm
DC}_{kk}$. The element  $W_{ij}$ of the binned Hamiltonian for bin size
$\Delta$ is then
\begin{equation}
   W_{ij} = \sum_{(i' - H_{kk})\leq \Delta,\; 
                   (j'- H_{ll})\leq \Delta}\; H_{kl}^2
  \label{binmat}
\end{equation}
where $i' = i\times\Delta $ and $j' = j\times\Delta$. The plot of binned 
Hamiltonian for Nd, Pm and Sm are shown in Fig.\ref{fighmat}. To improve the
contrast the plots show $\ln (W_{ij})$ instead of the $W_{ij}$. In all the
three atoms: (i) band like structure is evident, however, there are also
block structures within the band; (ii) another important feature is the 
prominent streaks of large matrix elements parallel to the diagonal, these
mix two diagonal blocks and,  as mentioned in \cite{gr-95}, perturbative
calculation can account for these distant elements; (iii) the Hamiltonian
matrices of Nd and Pm have small range of energy with no states and these 
are the empty stripes in the figures.

Presence of prominent diagonal blocks in the figures (point (i) above) is a
consequence of the spread of the various configurations shown in Fig. 4 and 
Table 1. Though it is tempting to conclude that there is a BRM like
structure, it should be noted that the various blocks retain their identity
each with their own internal variance $\sigma^2(\widetilde{m} \rightarrow
\widetilde{m})$. Fig. 9 clearly shows that the partial variances 
$\sigma^2(\widetilde{m} \rightarrow \widetilde{m}^\prime)$ (except in some
special cases) are in general much smaller compared to the internal
variances. It should be pointed out that block matrix structure as in Fig.
9 is also  observed very recently in  nuclear shell model calculations
\cite{Ze-96,papen-04} and it is plausible that such structure is a feature of
interacting particle systems. In \cite{Ze-96,papen-04} it is argued that one of
sources for this structure is $J$ symmetry. The various blocks are much more
well separated in Nd and Pm as compared to Sm and as a result Sm is much
closer to EGOE(1+2) with Nd and Pm exhibiting larger localization with
departures from EGOE(1+2) predictions.

To gain better understanding of the off-diagonal matrix (i.e. (i) and (ii)
above), which will mix the CSF's, let us consider their structure in some
detail. In the present calculations the off diagonal Hamiltonian matrix
elements  $H^{\rm DC}_{kk^\prime}$ are largely generated  by the
inter-electron Coulomb  interaction $1/r_{12}$ than the one electron terms
$t_i$ where, 
\begin{equation}
   t_i = c\bm{\alpha}_i\cdot\bm{p}_i + c^2(\beta_i - 1)-
         \frac{Z (\bm{r}_i)}{r_i}
  \label{one-body}
\end{equation} 
In second quantized notation, $H^{\rm DC}$ can be written as 
\begin{equation}
   H^{\rm DC} = \sum_{\alpha\beta}\langle\alpha | t|\beta\rangle 
                a^{\dagger}_{\beta} a_{\alpha} + 
                \sum_{\alpha\beta\gamma\delta}\langle\gamma\delta|
                \frac{1}{r_{12}}|\alpha\beta\rangle a^{\dagger}_{\gamma}
                a_{\delta}^{\dagger}a_{\beta}a_{\alpha}
\end{equation}
where for short we use Greek alphabets to denote orbitals and summation
is  without restrictions. The two-particle matrix element of $1/r_{12}$  can be
decomposed into multipole components, 
\begin{equation}
   \langle \gamma\delta|\frac{1}{r_{12}}|\alpha\beta\rangle = 
   \sum_K G^K(\gamma,\delta;\alpha,\beta) R^K(\gamma,\delta;\alpha,\beta)
  \label{r12sng}
\end{equation}
where $G^K(\gamma,\delta;\alpha,\beta)$  and 
$R^K(\gamma,\delta;\alpha,\beta)$ are angular and radial integrals 
respectively, and $K$ is the multipole. The angular momentum selection
rules  are consequence of $G^K(\gamma,\delta;\alpha,\beta)$ and the
radial integral  also known as Slater integral \cite{lindgren}.  In total
there are $N_v^2(N_c + N_v - 1)^2/4$ matrix elements of $1/r_{12}$ which
contribute to the off-diagonal Hamiltonian  matrix elements; $N_v$ and $N_c$
are defined in Section II. The many-particle matrix elements of $1/r_{12}$
can be written as, 
\begin{equation}
   \langle \gamma_rPJM|\frac{1}{r_{12}}|\gamma_sPJM\rangle = 
   \sum_{\alpha\beta\gamma\delta}\sum_K C_{rs}^{K, JM}(\gamma,\delta;\alpha,\beta) 
   R^K(\gamma,\delta;\alpha,\beta)
  \label{r12csf}
\end{equation}
where $C_{rs}^{K,JM}(\gamma,\delta;\alpha,\beta)$ is the angular factor which 
transports the single particle matrix elements to the many-particle space.
The $C$'s consists of two components, first the angular  factor in the single
electron space and second a component which propagate  from the single
electron to many electron space. These will have  complex structure when
electrons of several open-shells are coupled and the  distribution of the
non-zero Hamiltonian matrix exhibit intricate patterns.  Interestingly,
nuclear shell model calculations showed that \cite{Ze-96,papen-04} the angular
factors generate a part of the GOE structure of the diagonal blocks. The
radial integral is large  for $K=0$ in Eq. (\ref{r12sng})  and it is an allowed
multipole when $\kappa_{\alpha} = \kappa_{\gamma}$ and $\kappa_{\beta} =
\kappa_{\delta}$.  This can couple energetically  well separated
configurations strongly and explain (ii) above. Finally, from the matrix 
structure one can understand qualitatively the departures of $F_k(E)$ and
$\xi_2(E)$ from EGOE(1+2). In this random matrix ensemble the two-particle
matrix elements variance (the parameter $v^2$ in Eq. (\ref{v2var})) is 
independent of its position (i.e. independent of the indices 
$(i,j,k,l)$ in Eq. (\ref{v2var}) except that the diagonal matrix elements 
have twice the variance of the off-diagonal matrix elements).


\section{conclusions and future outlook}

In this paper wavefunctions of complex lanthanide atoms Nd, Pm and Sm are
analyzed in terms of strength functions, number of principal components,
occupation numbers and also the Hamiltonian matrix structure. Examination of 
$F_k(E)$ showed that the BW form is dominant in Nd, they are more towards
Gaussian in Pm and quite close to Gaussian in Sm for the basis states not
very far from the $\epsilon_k$ centroids. All three atoms exhibit BW form for
${\hat{\epsilon}}_k < -1$. Thus the statistical spectroscopy developed by
Flambaum et al \cite{Fl-96,Fl-97,Fl-97b} will be good for as long as $F_k(E)$
with ${\hat{\epsilon}}_k \lesssim -1$ contribute to observables of
interest. In all these atoms there are localized states throughout the
spectrum ( though the density is not high) and this is clearly seen in the 
measure NPC. The structure of the localized states is closely correlated with
the occupation numbers. All these and the matrix structures showed
that EGOE(1+2) need to be modified. One approach is to partition the
two-particle space and employ different variances for different block
matrices and then propagate the ensemble to many particle spaces. Very few
properties of these partitioned ensembles ($p-EGOE(1+2)$) are known till  now
\cite{Ko-01} and clearly they should be studied in more detail. The 
induced TBRE introduced recently \cite{Al-04} are closely related to p-EGOE(1+2). 
Another
approach is consider the two-particle Hamiltonian as a mixture of a regular
part (say determined by the lower order multipoles in Eq.(\ref{r12csf})) and a 
random part and analyze them as a function of the strength of the random part 
(they are called $K+\alpha EGOE(2)$ in \cite{Ko-01} where $K$ is a fixed 
operator, and numerically some studies of such an ensemble are 
carried out using nuclear shell model recently \cite{Nu-03}).
Thus an important outcome of the present detailed analysis (presented in
Sections II-V) is that for further progress in understanding the wavefunction
structure in complex atoms it is necessary to analyse various modified
EGOE's and this is for future.

\newpage


\newpage


\begin{figure}
   \epsfig{width = 3in, height = 7in, angle = -90, figure = 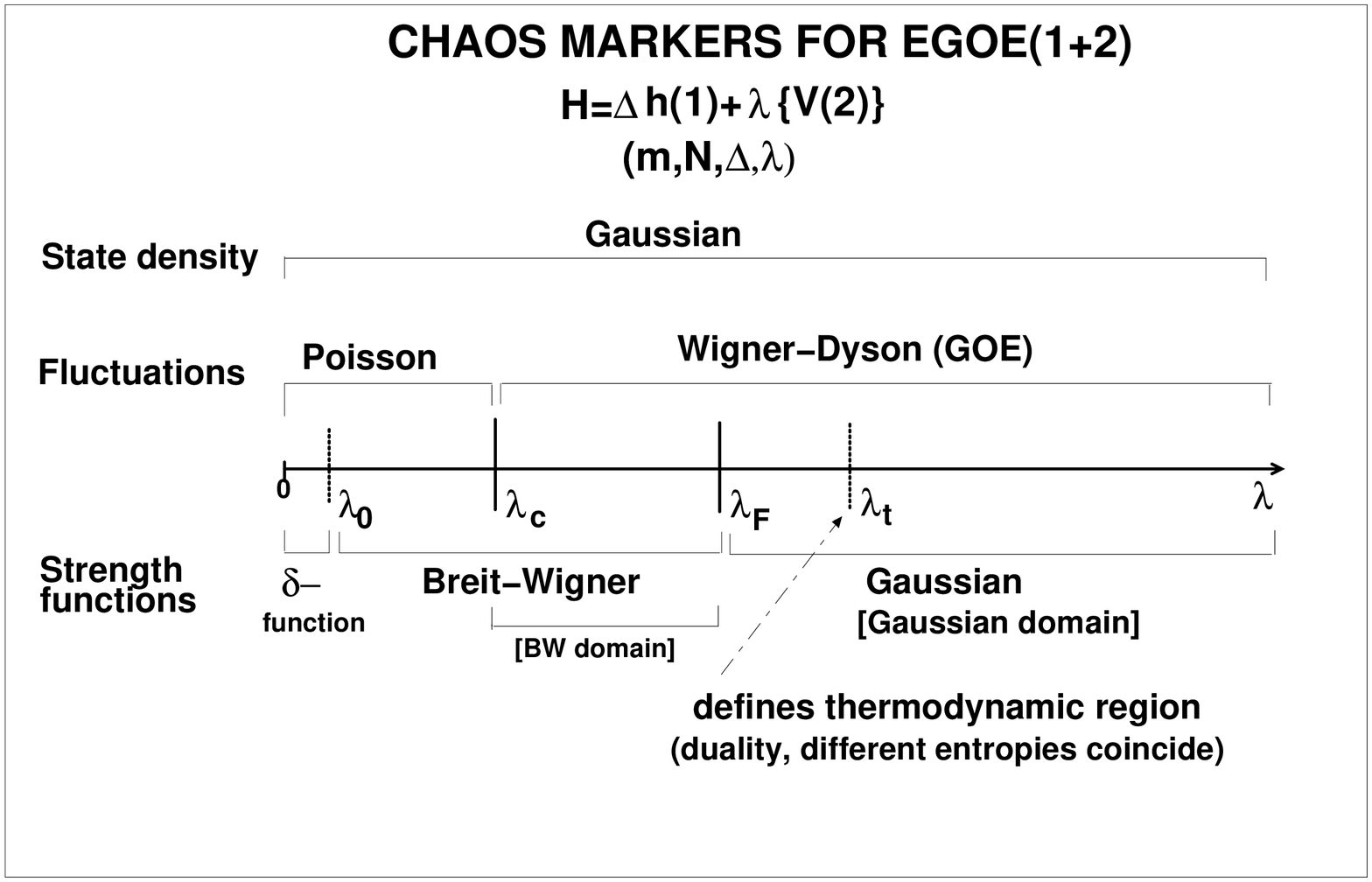}
   \caption{ Chaos markers for EGOE(1+2)
           }
  \label{cmark}
\end{figure}

\begin{figure}
   \epsfig{width = 4in, height = 3in, figure = 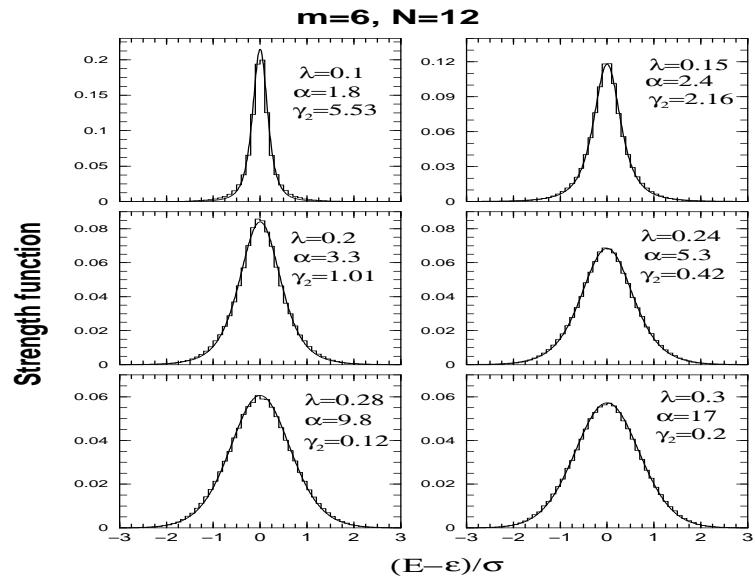}
   \caption{Strength functions for EGOE(1+2) as a function of the
            interaction strength $\lambda$. }
  \label{fke}
\end{figure}

\begin{figure}
   \epsfig{width = 6in, height = 3in, figure = 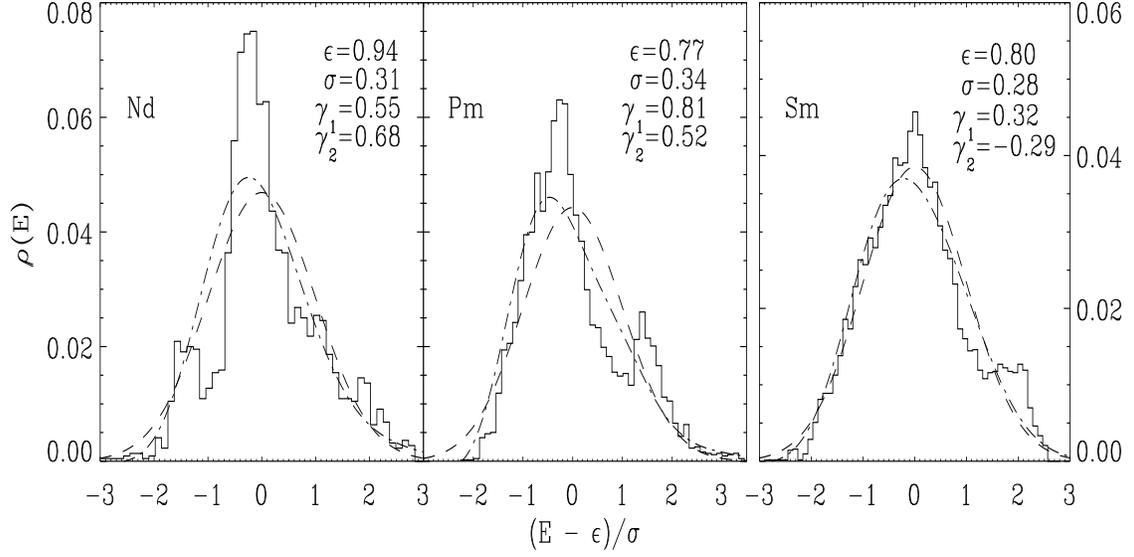}

   \caption{Binned density of states $\rho(E)$ of Nd, Pm and Sm. The
	    dashed  and dash-dot curves are the Gaussian and Edgeworth
	    corrected Gaussian representations. }
  \label{ndpmsm_den}
\end{figure}

\begin{figure}
   \epsfig{width = 6.5in, height = 3in, figure = 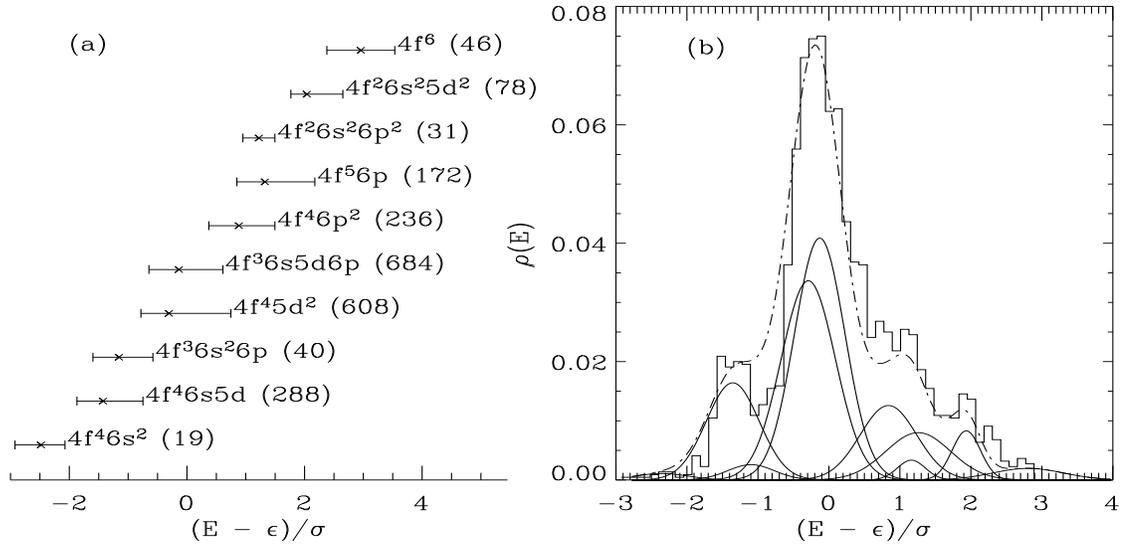}
   \caption{(a)The energy range of CSFs for Nd arising from a non-relativistic 
            configuration. The energies are calculated with respect
            to the  lowest CSF energy within the manifold 
            ( -9625.2394 hartree). The number of CSFs are given within 
            parentheses and the cross marks the location of the centroid.
            (b) The solid and dot-dash line are partial Gaussian densities 
            $\rho^{H, \widetilde{m}}(E)$ of each non-relativistic 
            configurations and sum of all the partial densities respectively. 
            The histogram is that of the density of states $\rho^H(E)$. } 
\label{ndcsf_engy} 
\end{figure}

\begin{figure}
   \epsfig{width = 7in, height = 6.5in, figure = 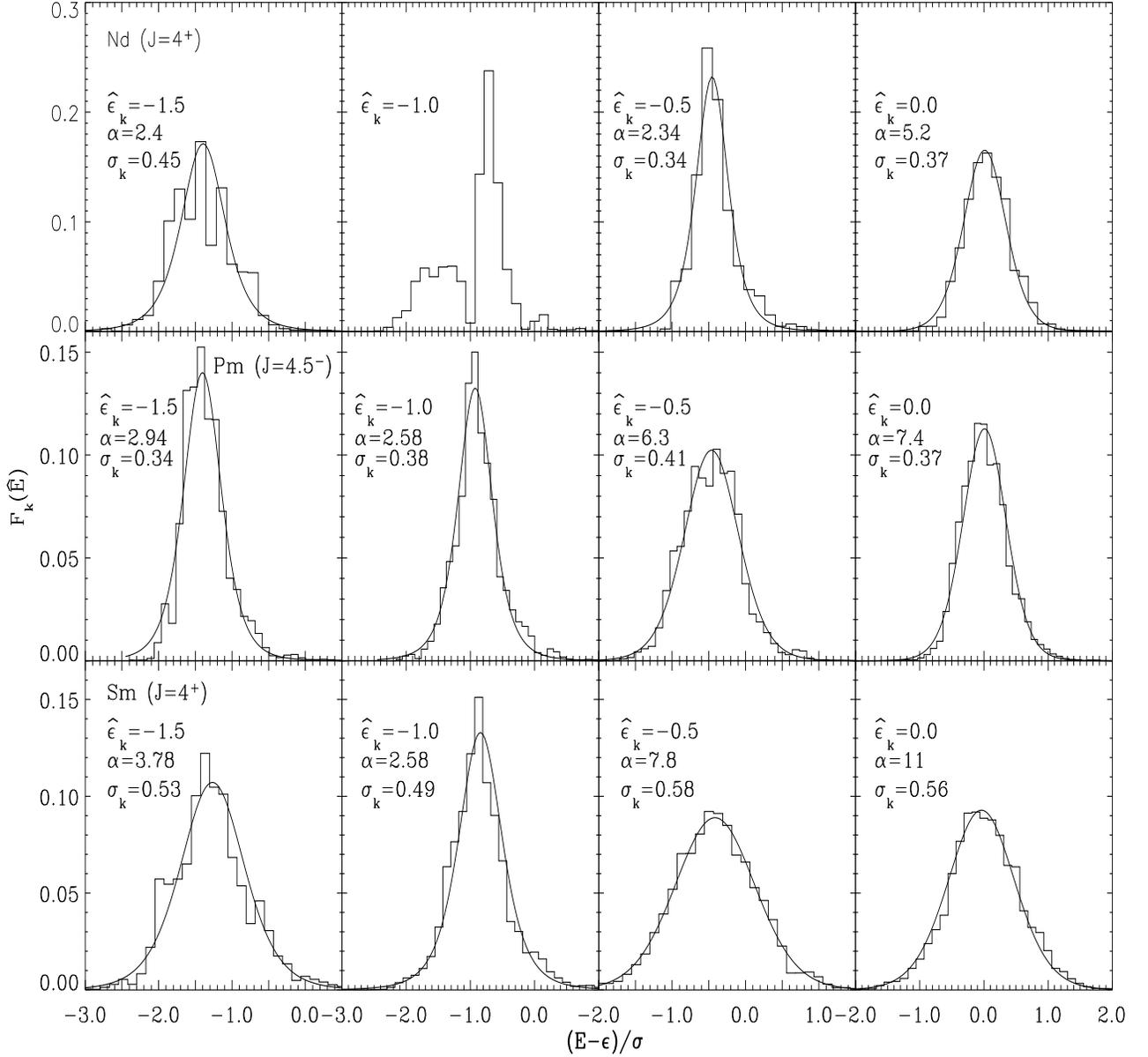}
   \caption{The plots in the three rows are the averaged 
	    strength function $\overline{F_k(\widehat{E})}$ for Nd, Pm and Sm
	    respectively. As labeled in the plots, the plots  in the four
	    columns show $F_k(\widehat{E})$ averaged in the range $-1.5
	    \pm 0.25\sigma _k$,  $-1.0 \pm 0.25\sigma _k$, $-0.5\pm
	    .25\sigma _k$,  and $0.0\pm .25\sigma _k$ respectively.}
  \label{ndpmsm_fk}
\end{figure}

\begin{figure}
   \epsfig{width = 7in, height = 6in, figure = 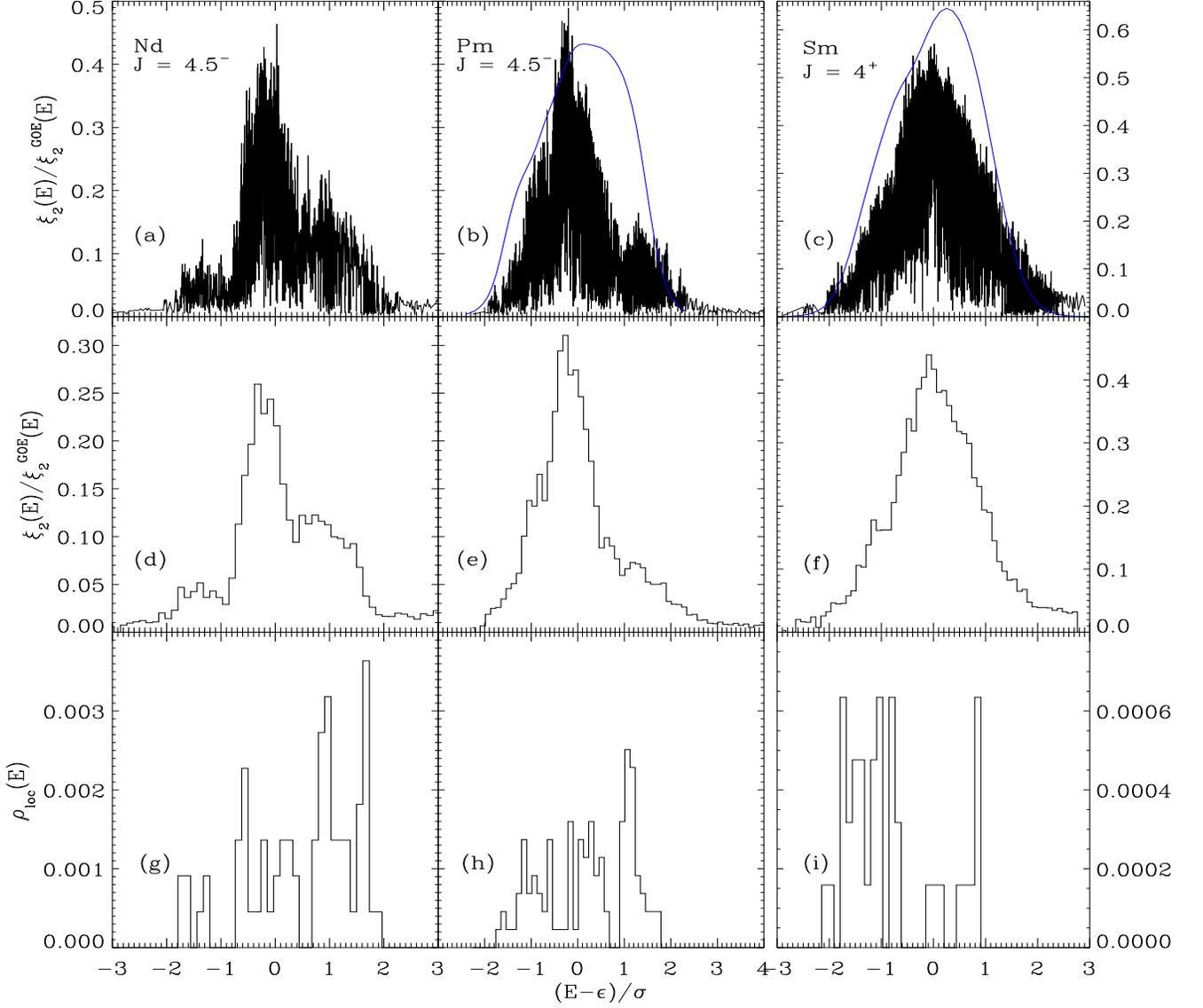}
   \caption{The number of principal components  $\xi_2(E)
	    =(\sum_k|C_k^E|^4)^{-1} $ of Nd I, Pm I and Sm I are plotted
	    in the first row. The plots in the second row  are binned
	    $\xi_2(E)$ and the last row is the binned density of localized
	    states $\rho_{\rm loc}(E)$. In (b) and (c) the continuous
            curves are from Eq. (\ref{xi2goe}) as described in the text.}

  \label{npcfig}
\end{figure} 

 \begin{figure}
   \begin{tabular}{cc}
   \epsfig{width = 6in, height = 8in, figure = 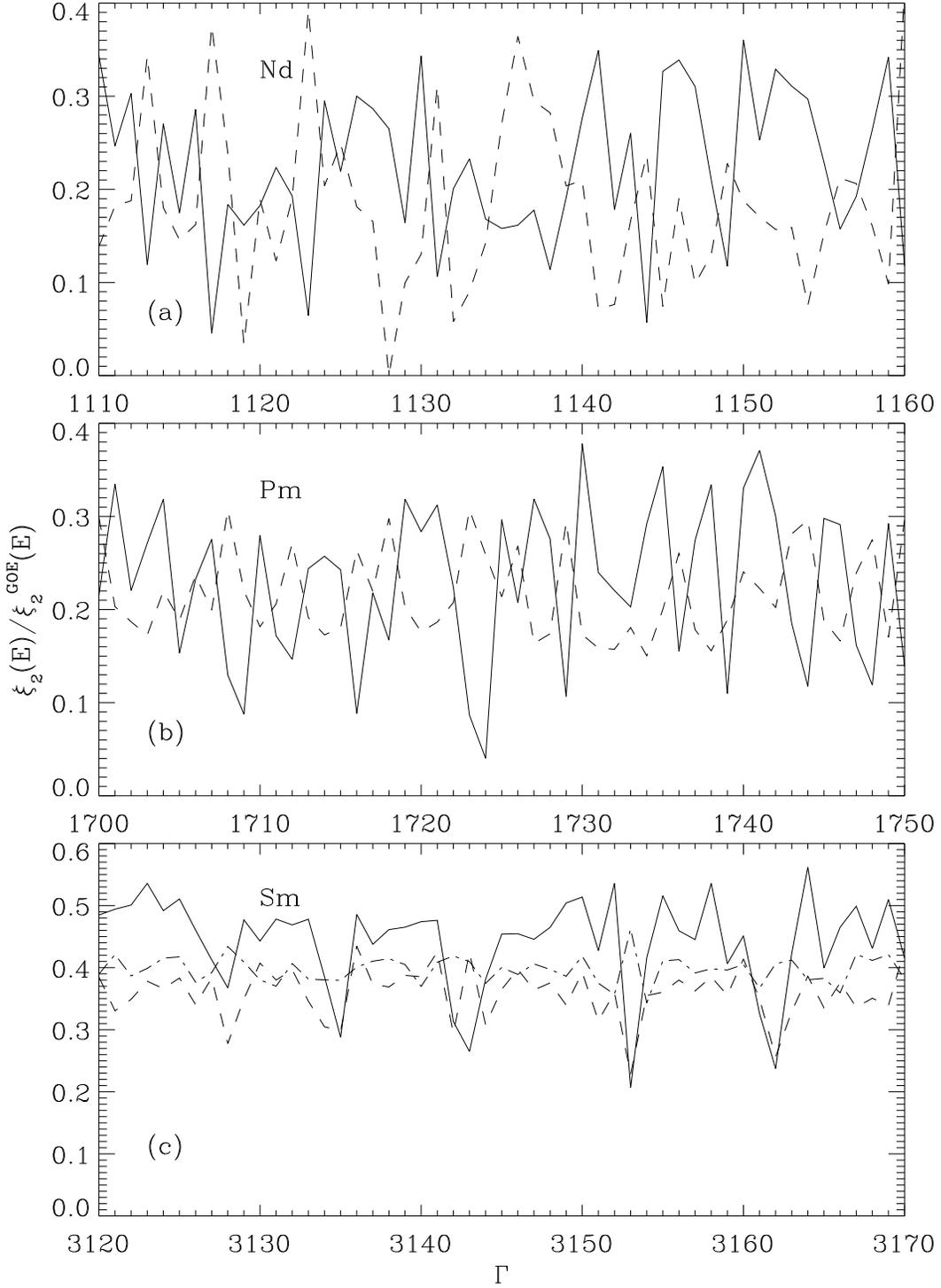} &
   \end{tabular}

   \caption{The solid line is the number of principal components $\xi_2(E)$ of 
            energy states $\Gamma$ around the centroid. The dot-dash line
	    show the occupation number $\langle n_\alpha\rangle^E$ of
	    selected shells which anti-correlate with the localized
	    states. The occupation numbers are scaled and shifted for
	    better comparison with $\xi_2(E)$: for Nd the plot shows
	    $\langle n_{5d_{3/2}}\rangle^E+\langle
	    n_{5d_{5/2}}\rangle^E-1$ , for Pm it is  $0.2(\langle
	    n_{5d_{3/2}}\rangle^E+\langle n_{5d_{5/2}}\rangle^E)$ and for
	    Sm the dot-dash and dashed lines are $\langle
	    n_{6s_{1/2}}\rangle^E-3$  and $\langle
	    n_{6p_{1/2}}\rangle^E+\langle n_{6p_{3/2}}\rangle^E-3$ 
	    respectively. }
 \label{occfig}
\end{figure}

\begin{figure}
   \epsfig{width = 6in, height = 8in, figure = 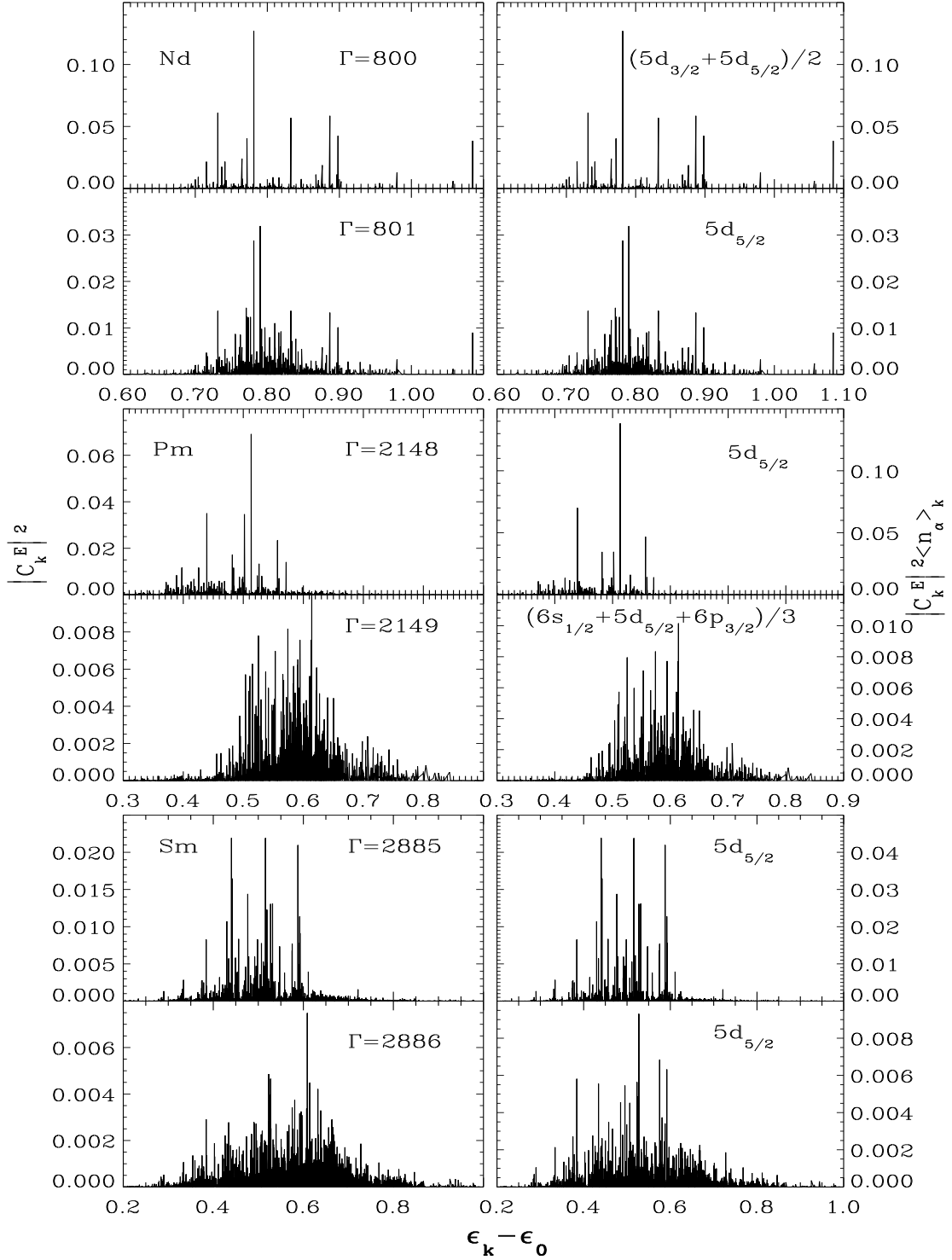}
   \caption{The $|C_k^E|^2$ of neighboring localized and chaotic pair of 
            $|E\rangle$ are shown in the plots in the first column. Even though 
            each pair are neighboring states, a large difference in the 
            structure is discernable.  For the localized states, the second
            column shows the individual $|k\rangle$ contribution 
            $|C_k^E|^2 \langle n_\alpha\rangle _k$ to the sub-shell of highest 
            $\langle n_\alpha\rangle$, whereas for chaotic states all the
            sub-shells have similar $|C_k^E|^2 \langle n_\alpha\rangle _k$ and 
            the plots show contribution from a randomly chosen sub-shell.
            }
  \label{egfn}
\end{figure}

\begin{figure}
   \epsfig{width = 3in, height = 7in, figure = 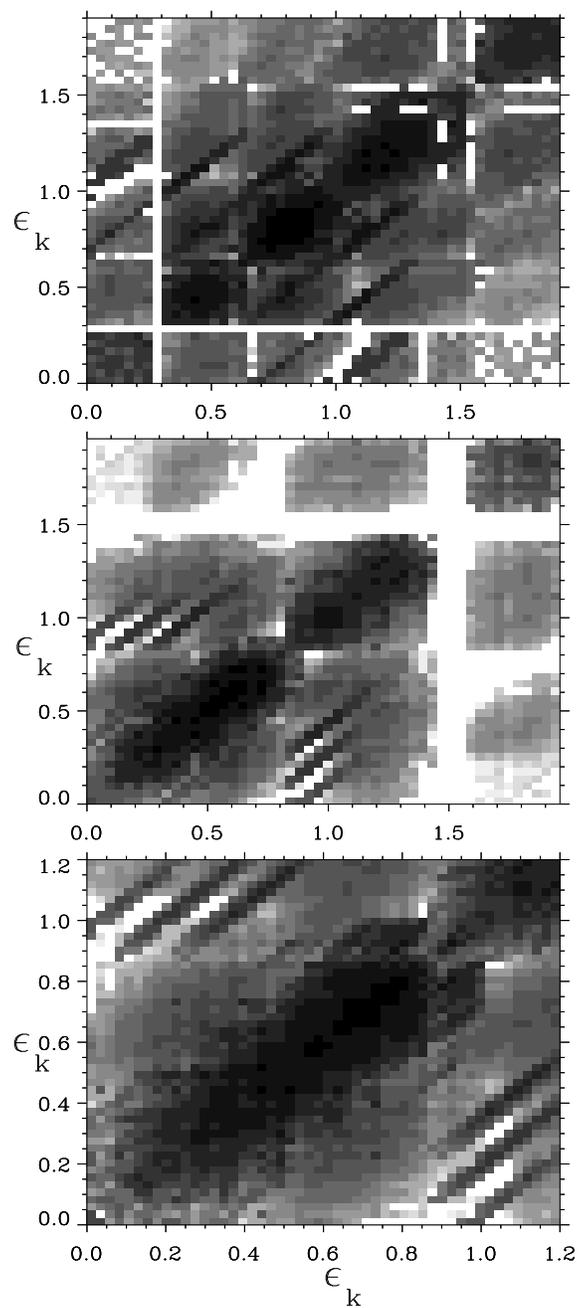}
   \caption{The binned Hamiltonian matrix of Nd I, Pm I and Sm I.
           }
  \label{fighmat}
\end{figure}

\end{document}